\documentclass[aps,onecolumn,preprintnumbers,nofootinbib,longbibliography,12pt,tightenlines]{revtex4-1}
\usepackage[utf8]{inputenc}
\usepackage{amsfonts,amssymb,stmaryrd,latexsym,amsmath,braket}
\usepackage{graphicx,subfigure}
\usepackage{times}
\usepackage{slashed}
\usepackage{color}
\usepackage{hyperref}
\usepackage[english]{babel}
\usepackage{amssymb,amsthm,amsmath,amstext,amsbsy,amsopn}
\usepackage{bbm}
\usepackage{nicefrac}
\usepackage{leftidx}
\usepackage{environ}
\usepackage{mathtools}
\usepackage{array}
\usepackage{isotope}
\usepackage{titlesec}
\usepackage{enumitem}

\makeatletter
\def\p@subsection{}
\def\p@subsubsection{}
\makeatother

\titleformat*{\section}{\LARGE\bfseries\sffamily}
\titleformat*{\subsection}{\Large\bfseries\sffamily}
\titleformat*{\subsubsection}{\large\bfseries\sffamily}

 \hypersetup{
     colorlinks=true,
     linkcolor=blue,
     filecolor=blue,
     citecolor = black,      
     urlcolor=cyan,
     }

\begin{document}

\pagenumbering{roman}

 \thispagestyle{empty} 
\includegraphics[width=\textwidth]{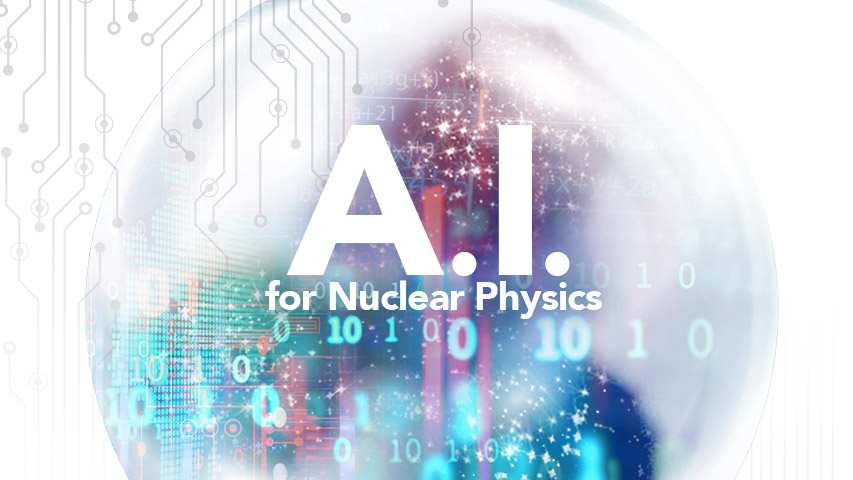}
\\ \\

\begin{center}
\begin{huge}
\textsf{This report is an outcome of the workshop \emph{AI for Nuclear Physics} held at Thomas Jefferson National Accelerator Facility
on March 4-6, 2020}  
\end{huge} 

\vfill 

\href{https://www.jlab.org/conference/AI2020}{\Large https://www.jlab.org/conference/AI2020}
\end{center}

\clearpage
\newpage
\includegraphics[width=\textwidth]{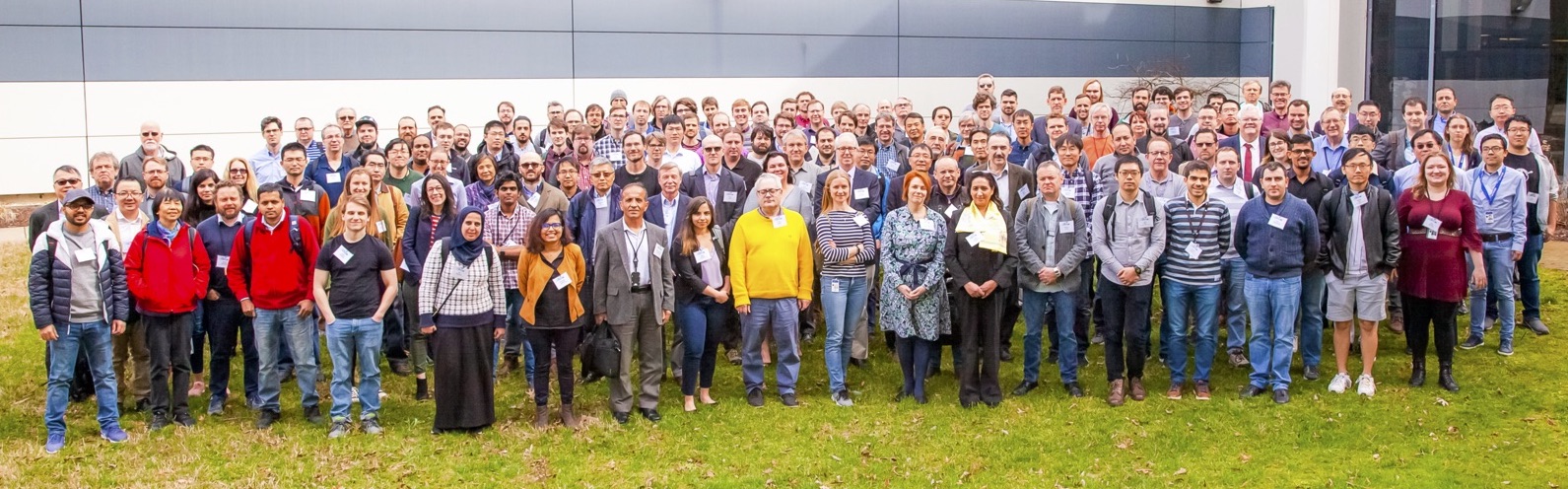}

\noindent 
Group photo from the workshop AI for Nuclear Physics held at Thomas Jefferson National Accelerator Facility
on March 4-6, 2020.

\vfill
\noindent
{\bf Disclaimer}\\
\noindent 
This report was prepared as an account of work sponsored by an agency of the United States Government. Neither the
United States Government nor any agency thereof, nor any of their employees, makes any warranty, express or implied,
or assumes any legal liability or responsibility for the accuracy, completeness, or usefulness of any information, apparatus,
product, or process disclosed, or represents that its use would not infringe privately owned rights. Reference herein to
any specific commercial product, process, or service by trade name, trademark, manufacturer, or otherwise, does not
necessarily constitute or imply its endorsement, recommendation, or favoring by the United States Government or any
agency thereof. The views and opinions of authors expressed herein do not necessarily state or reflect those of the
United States Government or any agency thereof.

\clearpage
\newpage

\noindent
{\Huge A.I. for Nuclear Physics}
\\ \\ 
\noindent
{\bf \large Editors}

\noindent 
{Paulo Bedaque}$^1$, 
{Amber Boehnlein}$^2$ (Lead Editor),
{Mario Cromaz}$^3$,
{Markus Diefenthaler}$^2$,
{Latifa Elouadrhiri}$^2$,
{Tanja Horn}$^4$,
{Michelle Kuchera}$^5$,
{David Lawrence}$^2$,
{Dean Lee}$^6$,
{Steven Lidia}$^6$,
{Robert McKeown}$^2$,
{Wally Melnitchouk}$^2$, 
{Witold Nazarewicz}$^6$,
{Kostas Orginos}$^{2,7}$,
{Yves Roblin}$^2$,
{Michael Scott Smith}$^{8}$,
{Malachi Schram}$^9$, and 
{Xin-Nian Wang}$^{10}$
\\ \\ 
$^1${\it University of Maryland}\\
$^2${\it Thomas Jefferson National Accelerator Facility}\\
$^3${\it Lawrence Berkeley National Laboratory}\\
$^4${\it Catholic University}\\
$^5${\it Davidson College}\\
$^6${\it Michigan State University}\\
$^7${\it College of William \& Mary}\\
$^8${\it Oak Ridge National Laboratory}\\
$^9${\it Pacific Northwest National Laboratory}\\
$^{10}${\it Lawrence Berkeley National Laboratory}

\vspace{20pt}
\hrule
\vspace{10pt}
\noindent 
This report is an outcome of the workshop \emph{AI for Nuclear Physics} held at Thomas Jefferson National Accelerator Facility
on March 4-6, 2020.
The workshop brought together 184 scientists to explore opportunities for Nuclear Physics in the area of Artificial Intelligence. The
workshop consisted of plenary talks, as well as six working groups. 
\vspace{10pt}
\hrule

\vfill

\noindent
This material is based upon work supported by the U.S. Department of Energy, Office of Science, Office of Nuclear Physics under contract DE-AC05-06OR23177.  Participation of students and early career professionals was supported by NSF, Division of Physics, under the grant ‘Artificial Intelligence (AI) Workshop in Nuclear Physics,’ Award Number 2017170.  Support for the Hackathon was provided by the University of Virginia School of Data Sciences and by Amazon Web Services

\clearpage\newpage

\tableofcontents

\setcounter{page}{3}
\pagenumbering{arabic}

\section{Executive Summary}

\noindent 
Nuclear science is concerned with the understanding of the nature of matter, its basic constituents and their interaction to form the elements and the properties we observe.
This includes the forms of matter we see around us and also exotic forms such as those that existed in the first moments after the Big Bang and that exist today inside neutron stars. The techniques, tools, and expertise needed for nuclear physics (NP) research are therefore diverse in nature. State-of-the art accelerators are being developed to illuminate the dynamical basis of
the core of the atom in terms of the fundamental constituents called quarks and gluons and to increase the number of isotopes with known properties.
This scientific infrastructure is reaching scales and complexities that require computational methods for tasks such as anomaly detection in operational data. New methodologies are needed to detect anomalies and to optimize operating parameters, 
predict failures as well as to discover new optimization algorithms.

Artificial Intelligence (AI)  is a rapidly developing field focused on computational technologies that can be trained, with data, to augment or automate human skill. Over the last few decades AI has become increasingly prominent in all sectors of everyday life, largely due to the adoption of statistical and probabilistic methods, the availability of large amounts of data, and increased computer processing power. 

The US government is initiating a broad-based, multidisciplinary, multi-agency program to build a sustained national AI ecosystem. Based upon two decades of research, development, and planning, the US government recognizes the importance of AI to advances in technology, national security and national infrastructure \cite{NAIRDStrategicPlan:2016}. The national AI Initiative \cite{WhiteHouseSummit:2019} provides a framework to establish a national strategy for US leadership in AI. Key areas of emphasis include: investments in AI research and development, unleashing AI data and resources, setting Government standards, and building the AI workforce. Several workshops and committees have identified the scientific opportunities for AI, as well as challenges from the intersection of AI with data-intensive science such as NP and high-performance computing. The present report is based on the “AI for Nuclear Physics Workshop” held in March of 2020  and outlines ongoing AI activities, possible contributions the NP community could make to identify and fill possible gaps in current AI technologies and needs to benefit NP research programs.
The Workshop brought together the communities directly using AI technologies and provided a venue for identifying the needs and commonalities. 

For the purpose of this report we define Artificial Intelligence (AI) to broadly represent the next generation of methods to build models from data and to use these models alone or in conjunction with simulation and scalable computing to advance scientific research. These methods include (but are not limited to) machine learning (ML)\footnote{Machine learning enables computers to learn from experience or examples.}, deep learning (DL)\footnote{Deep learning is a class of ML algorithm that are composed of multiple hidden layers.}, statistical methods, data analytics, and automated control. 

AI has tremendous potential within NP Research. It can provide new insights and discoveries from both experimental and computational data produced at user facilities. All top priorities of the 2015 Long-Range Plan on Research Opportunities and Directions \cite{Geesaman:2015fha} can benefit from AI. A common theme is to investigate and apply AI methods with well-understood uncertainty quantification, both systematic and statistical, to accelerator science, NP experimentation, and NP theory. At the same time, a number of activities and technologies in the diverse NP research portfolio has the potential to contribute to the emerging AI programs. For example, NP presents data on short time scales and with many different configurations that expose the limitations of current methods and could contribute to making AI more interpretable for the long term. 

A general characteristic of the application of AI in NP is the identification of small changes in patterns (statistical variations) in multi-dimensional and highly-correlated data (parameters, channels). This process includes the evaluation of models where one can use AI methods to identify the most promising computational pathways
where AI determined parameterizations can be used to avoid performance-limiting sections. Traditional AI tools have been applied successfully to some of these problems, in particular image classification. However, NP data are very diverse and to address the most interesting challenges more science insight has to be built into current AI technologies and AI tools have to be tuned to optimize performance in each application domain. Furthermore, NP data volume and complexity is increasing at a rapid pace.
To take full advantage of AI in NP will thus require investments and changes in methodology for the provisioning of computing and handling of data. This in turn will require adequate computing resources, e.g., access to GPU computing and disk storage at appropriate scales.

AI has the potential to transform NP. However, to fully realize AI contributions to NP, and vice versa, close collaboration among universities, technology companies, national laboratories, and other government agencies will be essential. Such collaboration will be required to bring, for example, state-of-the art AI techniques to the NP community. Workforce development is key to increase the level of AI-literacy in NP. The challenges are similar to those outlined in the NSAC Report on `Nuclear Physics and Quantum Information Science' \cite{QIS}  and include educational activities, creation of a community of AI knowledgeable researchers, and collaboration between NP and AI experts. Cross-disciplinary partnerships can help facilitate these connections. The list of Community identified Needs and Commonalities for AI Research essential for NP Applications as identified at this Workshop are presented below and also appear in more detail in Sec.~\ref{PRD} of the report:
\begin{enumerate}[label=(\roman*)]
\item Need for workforce development: There is a need to develop and sustain an AI capable workforce within NP.    
\begin{itemize}
\item Need for educational activities in AI: The goal is to retain talented students in AI-related fields and to help them to secure employment in a wide range of careers, thus ensuring that the new techniques and concepts developed in NP laboratories are widely disseminated. 
\item   Need for broader community: It is essential to have a community of researchers knowledgeable in AI technologies. 
\item   Need for collaborations: Long term commitment to partnerships between NP researchers and experts in AI/ML/Data Science is crucial as it takes time—for all parties involved—to learn the language and methods. 
\end{itemize}
\item   Need for uncertainty quantification:
The evaluation and comparison of uncertainty predictions using different modalities is required for widespread use of AI in NP.
\item Need for appropriate use of industry standard tools: significant effort is required in the careful tuning of ML tools (hyperparameter determination) to optimize performance in each application domain.  
\item   Need for problem-specific tools: the most interesting challenges that can be approached in NP with AI will require approaches that go beyond industry standard tools.  
\item Need for comprehensive data management: To maximize the usefulness of the data, it will be important to have standards on the processing of data, the application of theoretical assumptions, and the treatment of systematic uncertainties that will be used as training samples or as part of combined analysis. This meta-data will be encoded in the datasets.
\item   Need for adequate computing resources: AI techniques are computationally intensive and success in using these techniques will require access to GPU computing and disk storage at appropriate scales. 
\end{enumerate}

\section{Priority Research Directions}\label{PRD}
One aspect of the workshop was to explore areas where the application of AI could have a profound impact on Nuclear Physics Research.  This section summarizes those research directions.  Additional detail can be found in the Summary of Workshop Sessions.

\subsection{Future prospects}

\begin{description} 
\item[Accelerator design and operations] Many areas of accelerator design and operations will benefit from investments in AI and ML technologies. \\[-20pt]
\begin{itemize} 
\item 
Optimized design of accelerator systems. Development and validation of virtual diagnostics (e.g. longitudinal phase space monitors or predictors); Design and simulation of novel accelerators, and advanced engineered materials; Optimized diagnostic design and deployment; Improvement to beam sources and injector performance.
\item 
Improving facility performance and user experience. Data-driven beam generation, transport, delivery optimization; Automated learning for operator support; Hardware acceleration of ML in distributed control systems; Anomaly detection and mitigation (eg. LLRF, beam diagnostics); System health monitoring (e.g., targets, cryoplant); Data driven system maintenance.
\end{itemize} 
\item[Holistic approach to experimentation] As a long-term vision, disparate data sources (such as accelerator parameters, experimental controls, and the detector data) would be intelligently combined and interpreted to improve experiments. Real time analysis and feedback will enable the quick diagnostics and optimization of experimental setups. ML expert systems can increase the scientific output from the beam time allocated to each experiment. 
\item[Experiment design not limited by computation] Future experimental advances in accelerator-based NP research hinges on increased luminosity, which provides the statistics necessary to observe rare processes. ML methods will reduce computational barriers to reach this goal. Intelligent decisions about data reduction and storage are required to ensure the relevant physics is captured. Depending on the experiment, AI can improve the physics content of the data taken through data compactification, sophisticated triggers (both software and hardware-based), and fast-online analysis.
\item[Improving simulation and analysis] Improving simulation and data analysis using ML techniques is proceeding with two general aims: (i) to use these new techniques to improve the sensitivity of current instruments and accuracy of the data, and (ii) to decrease the time simulations and analyses  takes allowing for faster turnaround time to produce scientific results. Improving sensitivity allows more information to be extracted from datasets, which decreases uncertainty in results and increases discovery potential. Decreasing simulation and analysis time, saves costs and ultimately allows for a higher volume of scientific output by accelerating the feedback loop between experiment, analysis, and theory.

\item[Game changer in nuclear theory] A number of case studies have been identified. They are listed in the following.

\begin{description}[style=multiline,leftmargin=5cm,font=\normalfont]
\item[Sign problem in LQCD] The application of Monte Carlo techniques to systems at finite density (as in nuclear matter), real-time evolution (transport coefficients) and light-cone evolution (parton distribution functions) are hindered by the fermionic sign-problem. AI methods have begun to be applied, both in supervised and unsupervised learning modes. Potentially radical advances can be expected along this direction once the full power of AI is unleashed in this problem.

\item[Extraction of physical observables] To extract quantities of interest from correlation functions computed in LQCD in some cases requires the solution of an ill-defined inverse problem. AI methods now being applied to tackling the relevant inverse problems are showing great promise for achieving important milestones in our understanding of hadron structure from first principles.

\item[Propagator inversion in LQCD] The computation of observables in LQCD requires the calculation of quark propagators in the background of a large number of gauge configurations. Mathematically this requires the inversion of a large matrix. ML methods are beginning to be used to take cheaper inversions, done with low precision, and recovering the full precision propagator, with enormous savings in computer resources.

\item[Bayesian inference and global QCD analysis] Recent progress in ML with deep learning is enabling the development of new tools to advance the science of femtography, which shows great promise for high-precision determination of hadronic structure combining all available experimental data. Such approaches will be necessary for determining 3D nucleon tomography. 


\item[Identifying rare events] In the current approach to data taking and analysis, rare events, which can often represent major discoveries, can be easily overlooked when analysing data with preset ideas about what one is looking for. AI/ML can be used to generate events, with known theoretical parameters and models, and then compare the experimental stream readout with the pre-prepared theory expectations, to identify unusual or unexpected events that can be set aside for more focused study later.

\item[Microscopic description of nuclear fission] Various ML tools will help by dramatically speeding-up many-body simulations of nuclear fission by means of fast emulators for constrained density functional theory calculations in many-dimensional collective spaces; action minimization in the classically forbidden regions; new tools for dissipative dynamics; and computing of missing fission data.

\item[Origin of elements] A quantitative understanding of astrophysical processes responsible for the existence of elements requires knowledge of nuclear properties and reaction rates of thousands of rare isotopes, many of which cannot be reached experimentally. The missing nuclear data for astrophysical network simulations must be provided by massive extrapolations based on nuclear models. For some quantities such as nuclear masses, Bayesian ML has shown promise when aiming at informed predictions including both a reduction of extrapolation errors and quantified bounds.

\item[Quantified computations of heavy nuclei using realistic inter-nucleon forces] Predictions for heavy and very heavy nuclei such as Pb-208 using $A$-body approaches based on realistic two- and three-nucleon interactions with full uncertainty quantification will be enabled by Bayesian calibration using pseudo-data from microscopic calculations with supervised ML.

\item[Discovering correlations and emergent phenomena] Unsupervised learning can be used to discover correlations in nuclear wave functions based on microscopic Hamiltonians. There are terabytes of data from calculations with nucleonic degrees of freedom that can be data mined to discover emergent phenomena such as clustering, superfluidity, and nuclear collective modes such as rotations and vibrations.

\item[Development of a spectroscopic-quality nuclear energy density functional] Predictive and quantified nuclear energy density functional rooted in many-nucleon theory is needed. This development constitutes a massive inverse problem involving a variety of AI tools. The resulting spectroscopic-quality functional—crucial for understanding of rare isotopes—will properly extrapolate in mass, isospin, and angular momentum to provide predictions in the regions where data are not available.

\item[Neutron star and dense matter equation of state]  Data from intermediate-energy heavy-ion collisions and neutron-star merger events can be explored using AI tools to deduce the nuclear matter equation of state. ML classification tools can also be used in conjunction with calculations of infinite nucleonic matter to map out the phase diagram and associated order parameters.

\end{description}

\end{description} 

\subsection{Community identified Needs and Commonalities} 

\noindent
AI has tremendous potential within the context of NP Research.  However, the current AI tools and methodologies have limitations that have to be addressed for the long term.

\begin{description}

\item[Need for Workforce Development]
There is a need to develop and sustain an AI capable workforce within NP.  This challenge is similar to the workforce development challenge for Quantum Information Sciences, outlined in \href{https://science.osti.gov/-/media/np/pdf/Reports/NSAC_QIS_Report.pdf?la=en&hash=91703C70429F2B7D634CBC10573079858926141D}{`NP and Quantum Information Science’ Report}.

\textbf{\emph{Educational activities in AI:}} To this end, there is an urgent need to develop a range of outreach, recruitment, and educational activities. These activities will serve to raise interest in AI-related fields. The goal is to retain talented students in AI-related fields and to help them to secure employment in a wide range of careers, thus ensuring that the new techniques and concepts developed in NP laboratories are widely disseminated.
\\[-15pt] 
\begin{itemize} 
\item
University-wide AI courses: There is a need for inter-disciplinary AI courses involving Applied Mathematics, Statistics, and Computer Science experts, as well as domain scientists. 
\item
Graduate Fellowships are proven tools that enable the development of a well-educated workforce and could be used to good effect in the area of AI.
\end{itemize}
\textbf{\emph{Need for broader community:}}  To achieve the goals outlined by the community, it is essential to have a community of researchers knowledgeable in AI technologies. 
\begin{itemize}
\item
A centralized community based forum could provide a common foundation to build our technologies, allow for quick dissemination of new techniques, and provide a bridge from available AI resources to NP related applications.
\item
Successful inter-disciplinary research require mechanisms such as the ability to create joint faculty/staff appointments.  Given the wide range of use cases, such appointments would be beneficial at many institutions engaged in the NP Research Portfolio.
\end{itemize}

\textbf{\emph{Need for collaborations:}}  Collaboration with ML/AI/Data Science experts over a long-term is essential to successfully bring state of the art AI techniques to the NP community.   Long term commitment to partnerships between NP researchers and  experts in AI/ML/Data Science is crucial as it takes time -- for all parties involved -- to learn the language and methods.

\item[Need for problem-specific tools] The current surge in AI has provided great advances in software tools and hardware that can provide the basis of ML systems used in data processing. Readily available off the shelf solutions are well suited for several types of problems, particularly image classification. However, NP applications are unique in that they are often aimed at accelerating calculation, whether in the evaluation of models where one can use AI techniques to identify the most promising calculative pathways to simulation where AI-determined parametrizations can be used to circumvent performance-limiting elements.  While traditional ML tools may be applied to these problems, significant effort is required in the careful tuning of ML tools (hyperparameter determination) to optimize performance in each application domain.

\item[Enabling Infrastructure for AI in NP] 
Taking full advantage of AI for NP will require investments and changes in methodology for the provisioning of computing and handling of data.  Two particular areas concern data management and provisioning for resources.
\\ \\ [-10pt] 
{\bf Need for standardized frameworks:} The development of standardized frameworks such as ExaLearn and CANDLE have been extremely beneficial in other domains, and could provide a model for NP. It may be possible to adapt existing frameworks. 
\\ \\ [-10pt]
{\bf Need for comprehensive data management:} AI techniques are reliant on large volumes of data for training and the subsequent evaluation of models. For this reason, applications of AI are dependent on effective data management. Such data could be sourced from theoretical calculation, simulation, or experiment. Providing accessibility of the data to the wider NP community and increasing uniformity in data representation would create a connectivity across experiments that could increase collaboration and accelerate the development of AI techniques and tools. Such AI techniques could also facilitate near real-time calibration and analysis. To maximize the usefulness of the data, it will be important to have standards on the processing of data, the application of theoretical assumptions, and the treatment of systematic uncertainties that will be used as training samples or as part of combined analysis. This meta-data will be encoded in the datasets.\\ \\ [-10pt]  
{\bf Need for adequate computing resources} AI techniques are computationally intensive and success in using these techniques will require access to GPU computing and disk storage at appropriate scales. 

\item[Need for uncertainty quantification] A common theme is to investigate and apply AI methods with well-understood uncertainty quantification, both systematic and statistical, to accelerator science, NP experimentation, and NP theory. The commonly used ML algorithms do not provide error estimations with model predictions, which are essential to understand outcomes. In addition, an evaluation of metrics for the evaluation and comparison of uncertainty predictions using different modalities is required for widespread use of AI in NP.

\end{description}

\section{Workshop Overview}

The AI for Nuclear Physics Workshop was held at Thomas Jefferson National Accelerator Facility March 4-6, 2020.  The intent of the workshop was to make a broad survey of current AI projects in NP and to gather community driven input towards establishing priority research directions, areas of commonality across the NP community (and beyond), and general needs, including workforce development.  The agenda  focused on plenary sessions in the morning with topical working sessions in the afternoon, with most of the presentations available from the agenda.   184 people attended the workshop. The \href{https://www.jlab.org/indico/event/357/overview}{AI for Nuclear Physics Workshop Agenda} focused on summaries of status of the usage of AI in Nuclear Theory, Nuclear Experiment and Accelerator Science and operations. The connection between the scientific goals outlined in the Nuclear Science Advisory Committee long range plan \cite{Geesaman:2015fha} and AI was presented by Tim Hallman, Department of Energy Associate Director for the Office of Nuclear Physics. A second focus was the connection to broader efforts within DOE, including overview talks from the DOE Artificial Intelligence Technology Office, a summary of the AI for Science Townhall process, and a summary of the NeuroData without Borders Project \cite{NeuroData} and the Exascale Computing Project applications ExaLearn and CANDLE \cite{CANDLE} projects.       

An adjunct hackathon event was held on March 3, 2020.   8 teams each with four members participated.  The challenge problem was drawn from a common task in NP, measuring the properties of charged particles traversing a detector.  The challenge was structured as progressive, with five sub-challenges.    To enable evaluation of the success of the teams, an automated scoring system and leader board was developed, with the top two scoring teams being awarded prizes.  The computational approaches and tools used by the teams had significant variation, demonstrating that creativity in problem solving remains a feature of research undertaken with AI  Events such as this can be useful for furthering skills in AI for  participants who already have basic knowledge. 

\section{Summary of Workshop Sessions}
To serve as a record of the discussions, the conveners of the working group sessions have prepared summaries based on the workshop discussions and presentations.  The discussions reflect independent deliberations, and consequently some differences of opinion.  A list of the working groups and conveners are listed in the Appendix~B. As a note, due to conflicting workshops, some NP communities were not properly represented at this workshop.  Where possible, contributions from those communities were solicited and appear at the end of this section. 

\subsection{Lattice QCD and Other Quantum Field Theories}
Lattice field theory is a cornerstone of all subfields of NP, from nuclear structure to hadronic physics, heavy-ion collisions, and neutron stars. It is based on the Monte Carlo evaluation, in one guise or another, of the quantum path integral. Despite enormous successes achieved in the last few years, computing power currently  prevent us from addressing many of the central questions of NP. 

Lattice calculations are divided into the generation of gauge configurations, calculation of the observables of interest and data analysis. Artificial intelligence techniques have begun to be applied to all these stages as well as extending the applicability regime of lattice techniques.

\subsubsection{Case Studies and Future Prospects} 

{\bf Sign problem: }  The application of Monte Carlo techniques to systems at finite density (as in nuclear/neutron matter), real-time evolution (transport coefficients) and light-cone evolution (parton distribution functions) are hindered by the famous sign-problem.  It has been realized recently that the sign problem can be solved or ameliorated by evaluating the path integral not over real fields but over a manifold deformed into complex space instead. Up to now, the choice of manifolds has been guided by either impossibly expensive calculations or (human) insight into particular models. AI methods have begun to be applied, both in supervised and unsupervised learning modes \cite{Alexandru:2018ddf,Mori_2017,Alexandru:2018fqp,Bursa:2018ykf,Ohnishi:2019ljc}.
Potentially radical advances can be expected along this direction once the full power of AI is unleashed in this problem.

{\bf Configuration generation:} The usually most expensive part of a Monte Carlo calculation is the generation of configurations through the use of a Markov chain where, at each step, a new configuration is proposed and accepted or rejected with a probability depending on the new and old configurations. The practical feasibility of the method relies on being able to propose configurations that are significantly different from the old one while at the same time are likely to be accepted. The method used almost universally in QCD is the hybrid Monte Carlo algorithm (invented by the lattice QCD community and now widely used in all branches of science) becomes extremely expensive as the continuum limit is approached. A significant effort is being put into using different AI techniques to create algorithms to make better proposals, more decorrelated and more ``acceptable",  in order to speed up the process \cite{Albergo:2019eim,Kanwar:2020xzo}. The training of the algorithms is accomplished either by the use of configurations generated by standard algorithms or, more ambitiously, through fully unsupervised learning.
The basic ideas of such algorithms are already developed by the AI community and used for various applications in the engineering and software industry.

{\bf Propagator inversion: }
 The computation of observables in lattice QCD requires the calculation of quark propagators in the background of a large number of gauge configurations. Mathematically this requires the inversion of a large matrix and, in some applications, like the extraction of nuclear forces,  it can be the most expensive part of the calculation. Machine learning methods are beginning to be used to take cheaper inversions~\cite{Pederiva:2020}, done with low precision, and recovering the full precision propagator, with enormous savings in computer resources.
 
  {\bf Observables: }The extraction of physical observables from correlation functions computed in lattice QCD in some cases requires the solution of an ill-defined inverse problem. Such problems include the computation of parton distribution functions, generalized parton distribution functions, and transverse momentum dependent distribution functions, as well as the extraction of spectral densities and scattering phase shifts. These observables are the prime objective of the JLab 12GeV program where the 3D structure and spectrum of hadrons are studied, as well as the heavy-ion physics community. AI methods are now being applied to tackling the relevant inverse problems to address these physics goals showing great promise for achieving important milestones in our understanding of hadron structure from first principles.

\subsubsection{Enabling Discoveries/What is Needed}

All the work summarized in this section is exploratory.  The potential is enormous although at this time, the AI techniques are not yet competitive with the standard in the field numerical investigations of quantum field theories. That said, success in a single one of the approaches has revolutionary potential in the field. The approach for AI studies is based on toy models and small lattices where novel ideas can go through the cycle of implementation/testing/improvement very quickly. This requires a model of support that favors small, flexible groups, fosters informal communication between researchers both within NP and the AI community while keeping the field attractive to young people who may have options to pursue a career in the private sector.


\subsection{Low-Energy Nuclear Theory}
\subsubsection{Current Status}

ML applications of layered feed-forward networks  to modeling nuclear masses and other observables were carried out in the early 1990s \cite{Gazula1992,Gernoth1993}. But it is only fairly recently that the AI tools have been more broadly adopted by nuclear theorists and applied to various problems in nuclear structure and reactions.  The main areas of modern AI applications are the following: fast emulation for big simulations;
revealing the information content of measured observables with respect to current theory;
identifying crucial experimental data for better constraining theory;
revealing the structure of theoretical  models by means of advanced parameter estimation  and model reduction;
uncertainty quantification of theoretical results;  and
improving the  predictive capability by assessing extrapolations,
as theoretical models are often applied to entirely new nuclear systems and conditions that are not accessible to experiment.
\begin{figure}[htb]
\includegraphics[width=0.6\linewidth]{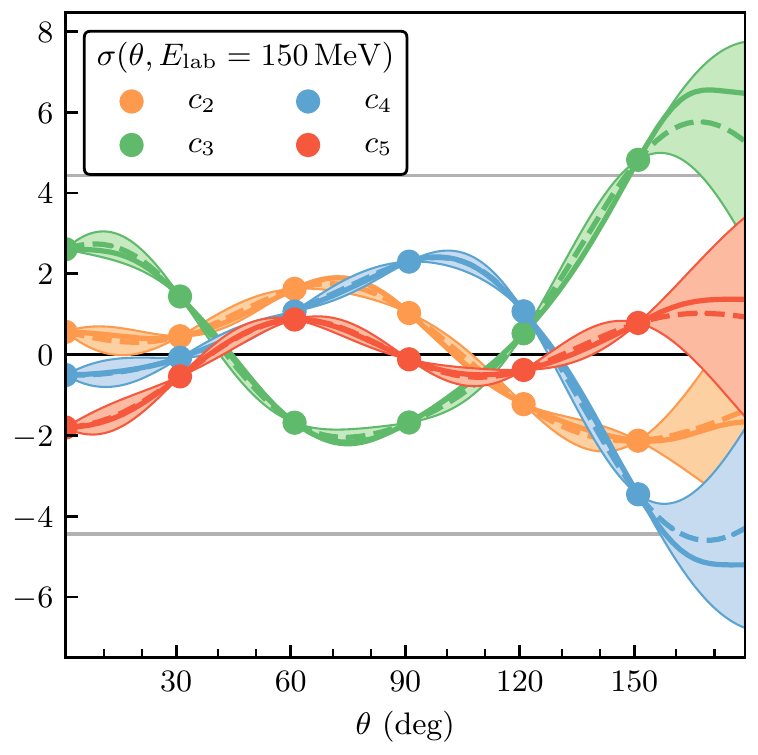}
\caption{{\bf Bayesian calibration.} Effective field theory analysis of neutron-proton scattering cross section using Bayesian Gaussian processes \cite{Melendez:2019izc}. The use of Gaussian processes permits efficient and accurate assessment of credible intervals.}
\label{NNGP}
\end{figure}

\begin{figure}[htb]
\includegraphics[width=0.8\linewidth]{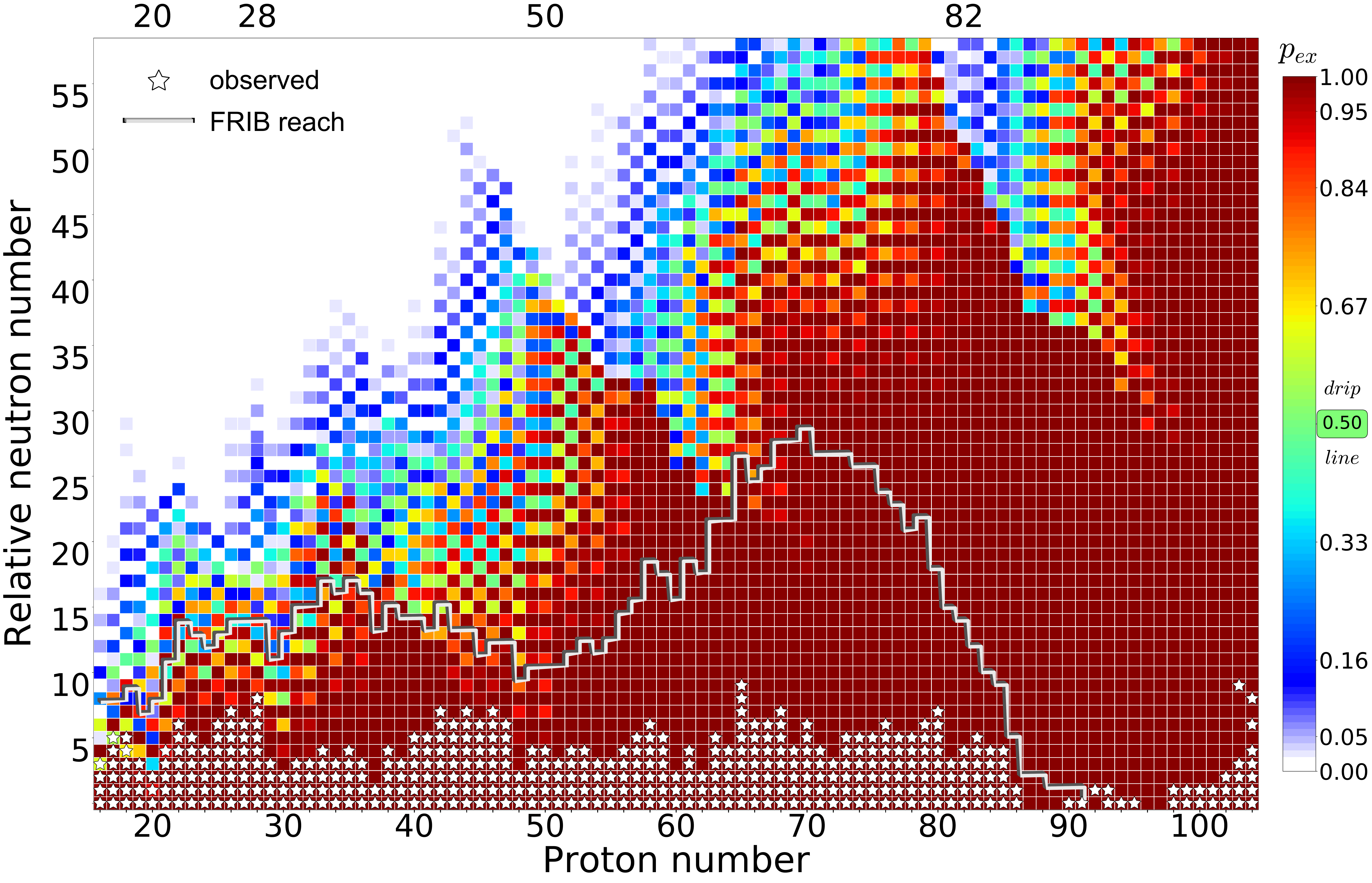}
\caption{{\bf Bayesian extrapolation and model averaging.} The quantified  separation energy landscape in the neutron drip-line region obtained with the Bayesian model averaging \cite{Neufcourt2020a}.
The color marks the probability  $p_{\rm ex}$ that  a given isotope is  bound with respect to  neutron decay. 
For each proton number, $p_{\rm ex}$ is shown  along the isotopic chain versus
 the  neutron number relative to that of the heaviest isotope
 for which a neutron separation energy  has been measured.
The domain of nuclei that have been experimentally observed is marked by stars. 
To provide a realistic estimate of the discovery potential with  modern radioactive ion-beam facilities,
the isotopes within  FRIB's experimental reach are delimited by the shadowed solid line.}
\label{nrich}
\end{figure}

A variety of AI/ML tools have been used: various flavors of neural networks, Bayesian calibration, Bayesian model averaging, radial basis function, and support for vector machines. The application areas include interpolation and extrapolation of
nuclear masses \cite{Utama16,Utama17,Niu2018,Neufcourt2018,Neufcourt2019a,Neufcourt2020,Neufcourt2020a,Niu2019,Sprouse2019,Pastore2019}, charge radii \cite{Utama16bis,Ma2020}, excited states \cite{Akkoyun:2020yfw,Akkoyun:2020nxa,Regnier:2019fgf}, beta decay
 \cite{Niu:2018trk,Costiris:2008yw}, alpha decay \cite{Rodriguez:2019rnj,BanosRodriguez:2019qgm}, fission yields \cite{Wang2019,Lovell2019},
 nucleon-nucleon phase shifts \cite{Wesolowski:2018lzj}, scattering in the unitary limit \cite{Kaspschak:2020ezh}, neutron-alpha scattering and three-body parameters \cite{Kravvaris:2020lhp}, nuclear reactions cross sections \cite{King:2019sax,Lovell:2018bao,Catacora-Rios:2019goa,Ma2020r,Akkoyun2020r}, estimates of
 truncation errors  \cite{Melendez:2017phj,Melendez:2019izc,Jiang2019,Negoita2019},   model calibration and reduction \cite{McDonnell2015,Yoshida2018,Yoshida:2019asd,Ekstrom2019,Ekstrom2019a,Ekstrom:2019hlw,Kejzlar:2020vla}, and variational calculations \cite{Keeble:2019bkv}.

 The low-energy nuclear theory community has been involved in educational efforts in the area of AI. Examples are \href{https://indico.frib.msu.edu/event/16/timetable/?view=standard}{summer schools}, \href{https://fribtheoryalliance.org/TALENT/courses/course_11.php}{courses}, and conferences, including a series of annual meetings on enhancing the interaction between nuclear experiment and theory through information and statistics (\href{https://isnet-series.github.io/meetings/isnet/}{ISNET}).

\subsubsection{Case Studies and Future Prospects} 

The following case studies are examples of high-impact science that can be
enabled by AI.
\noindent
{\bf Microscopic description of fission.}
Modern many-body approaches to fission \cite{schunck2016}, aided by AI, will provide a predictive description of fission that will produce   data for  heavy-element research, nuclear astrophysics, and stockpile stewardship. Here, AI-tools will help on several levels, including: development of emulators for constrained density functional theory  calculations in many-dimensional collective spaces \cite{McDonnell2015,Akkoyun2013},  action minimization in the classically forbidden regions, description of dissipative dynamics, and the use of neural networks to compute incomplete fission data \cite{Wang2019,Lovell2019}.

\noindent
{\bf Origin of heavy elements.}
The astrophysical rapid neutron capture r-process responsible for the existence of many heavy elements is predicted to involve many elements that are close to the  neutron drip line; the structure of these very exotic nuclei thus directly impacts how elements are produced in stellar nucleosynthesis
\cite{Horowitz2019}.
A quantitative understanding of the r-process requires knowledge of 
nuclear properties and reaction rates of $\sim$3,000  very neutron-rich isotopes, many of which cannot be reached experimentally. The missing nuclear data for astrophysical simulations must be provided by massive extrapolations based on nuclear models augmented by the most recent experimental data. Here, ML, with its unified statistical treatment of all uncertainties, can make informed predictions for some of the relevant quantities that reduce extrapolation errors and quantified bounds \cite{Utama17,Neufcourt2018,Neufcourt2019a,Sprouse2019}.

\noindent
{\bf Quantified computations of heavy nuclei using realistic inter-nucleon
forces.}
Predictions for heavy and very heavy nuclei such as $^{208}$Pb using $A$-body approaches based on realistic two- and three-nucleon interactions with full uncertainty quantification will be enabled by Bayesian calibration
using pseudo-data from microscopic calculations with supervised ML \cite{Ekstrom2019,Ekstrom2019a}.

\noindent
{\bf Development of a spectroscopic-quality nuclear energy density
functional:}
Predictive and quantified nuclear energy density functional rooted in many-nucleon theory \cite{Navarro2018} will be developed. This task constitutes a massive inverse problem 
\cite{Kortelainen2014} involving  a variety of AI tools.
The resulting spectroscopic-quality  functional---crucial for
understanding of rare isotopes---will properly extrapolate in mass, isospin, and angular momentum to provide predictions in the regions where data are not available.

\noindent
{\bf Discovering nucleonic correlations and emergent phenomena.}
Unsupervised learning can be used to discover correlations in calculations of nuclear wave functions that use a microscopic Hamiltonian. There are terabytes of data
from calculations with nucleonic degrees of freedom that can be data mined
to discover emergent phenomena such as clustering \cite{Elhatisari:2016owd,Elhatisari:2017eno,Freer:2017gip,Dawkins:2019vcr}, superfluidity \cite{Sambataro:2019jcp}, and nuclear rotation \cite{Caprio:2019yxh}.

\noindent
{\bf Neutron star and dense matter equation of state}
Data from intermediate-energy heavy-ion collisions and  neutron-star merger events can be explored using AI tools to deduce
the nuclear matter equation of state~\cite{Morfouace2019,Tsang2019,Lim2019}.
ML classification tools can also
be used in conjunction with calculations of infinite nucleonic matter to map out the phase diagram and associated order parameters.

\subsubsection{Enabling Discoveries/What is Needed}
The low-energy nuclear theory community is eager to embrace the diverse toolbox  offered by AI. Progress in the field could be accelerated by deploying  additional resources to meet the most important needs. 

\noindent
{\bf Need for  collaborations.} Many barriers can be overcome by establishing collaborations that have long-term perspective.
Considering the low level of AI literacy  in the community, access to ML/AI/Data science experts is essential. 
(Semi-)Permanent access to experts in AI/ML/Data Science is crucial as it  takes time---for all parties involved---to learn the language and methods. The best solution is to hire a AI/ML/Data Science expert as a joint faculty (or postdoc). 
Funding mechanisms should be defined to support local and national collaborations in NP and ML/AI/Data science.

\noindent 
{\bf Need for inter-disciplinary research.} Inter-disciplinary research is popular but making it succeed is difficult. Disciplinary boundaries mitigate against hiring ML/AI/Data Science experts involved in NP research. The silo mentality, especially in academia,   is a serious problem and is hurting innovation. Formal mechanisms must address the issues of how scholarship is assessed and how  teaching is assigned and evaluated, particularly before tenure.
Programs should be established  to fund AI/NP bridge positions at universities; this would help to create joint faculty appointments at many institutions.

\noindent 
{\bf Need for a comprehensive approach to AI education.} There is, at present, only a patchwork of AI educational efforts in the low-energy nuclear-physics community.
 A coherent  approach to AI education, involving multiple university departments, such as   Physics, Statistics, and Computer Science, is needed. While online courses can be effective, they cannot replace regular in-person lectures.
Establishing graduate fellowships in the area of ML/AI/Data science applied to NP problems would enable the development of a well-educated workforce in this area. Some universities have ``dual Ph.D.'' programs that allow individual students to work within two different graduate programs. Certificates in AI/ML are a less intensive but still beneficial approach to this problem.

\subsection{Accelerator Science and Operations}
We identify three distinct areas where AI/ML  could improve the reliability and performance of the NP accelerator facilities while reducing the operational cost. These areas are:
\begin{itemize}
\item Accelerator and material design optimization
\item Provenance and prognostication for accelerator sub-systems   
\item Dynamic optimization of real time operation controls
\end{itemize}
Although these areas can be investigated independently, providing a "optimal automated accelerator" would require all areas. 

\subsubsection{Accelerator and material design optimization}

Computational techniques lay at the center of accelerator design. Modern simulation codes are capable of self-consistent tracking  \(10^{9}\) charged particles through complex, nonlinear external field environments, and in modeling interactions with materials. Highly developed and benchmarked engineering codes are employed to design and optimize acceleration structures, high power beam targets, vacuum systems, plasma and solid-state devices for instrumentation.\\

ML/AI techniques are coming into common use during the design stage to facilitate studies of complex beam dynamics in search of optimum lattices and working point tunes, to study novel schemes for cooling hadron beams, to improve diagnostic schemes for beam measurements, to create performance gains in high intensity and high brightness beam sources, to name but a few \cite{Edelen:2019}. \\

Reinforcement learning and Bayesian optimization are techniques that can be used to explore large design parameter spaces. However, in order for these techniques to provide reliable and optimal solutions they need to be configured and tuned for the specific application. An incorrect kernel selection used in a Gaussian Process technique can lead to disastrous results. Similarly, using a sub-optimal search strategy and/or policy model architecture in reinforcement learning will converge to sub-optimal result. Therefore, it's critical to build or leverage a framework, such as CANDLE and ExaRL, to improve the chances of an optimal solution.

\subsubsection{Provenance and prognostication for accelerator sub-systems}

\begin{figure} 
    \includegraphics[scale=0.5]{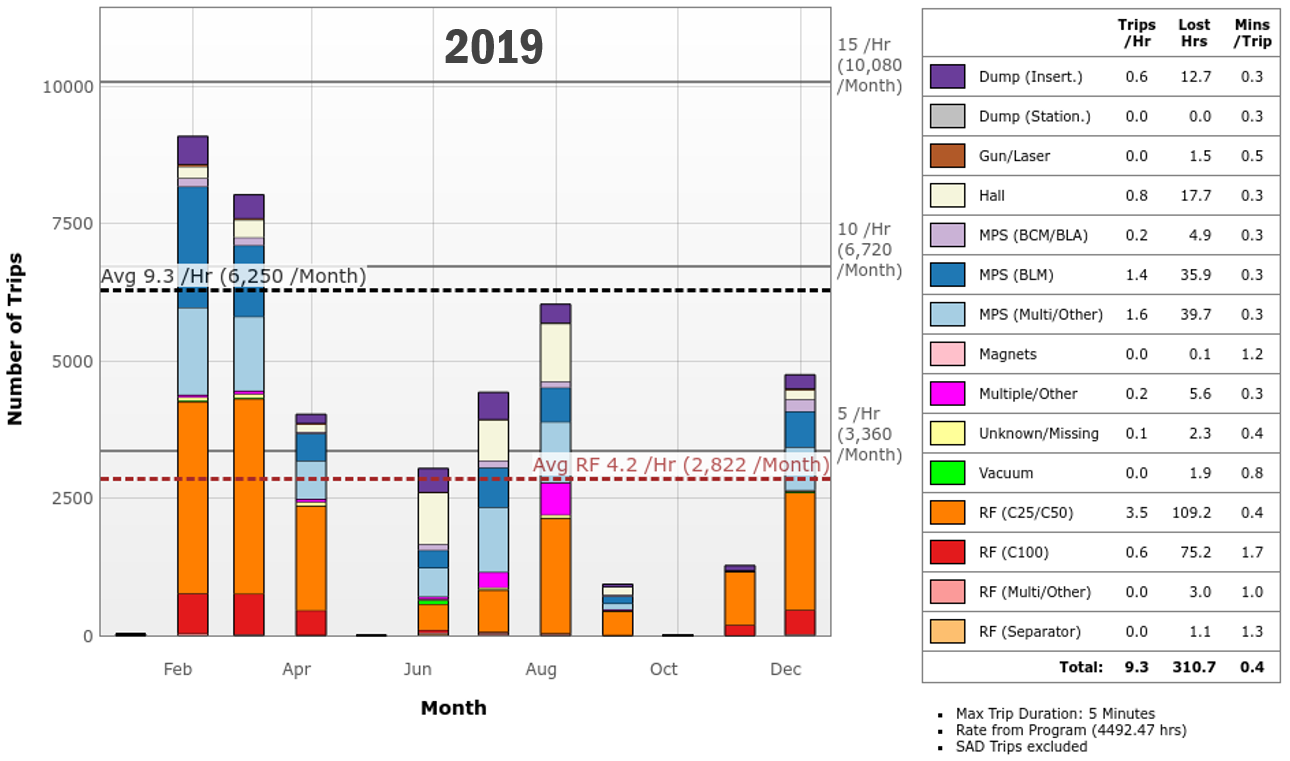},
    \caption{CEBAF beam trip event rates (Courtesy R. Michaud, Jefferson Lab).}
    \label{fig:CEBAF_trip_rates}    
\end{figure}

Scientific productivity at accelerator-based NP facilities is directly impacted by unscheduled losses of beam time. The trip rate (see Figure~\ref{fig:CEBAF_trip_rates}) is attributable to multiple causation factors that vary in frequency and severity. Some of the main causes are due to excessive beam losses detected by the Machine Protection System (MPS) and to loss of RF cavity control (RF). Machine learning tools for anomaly detection have been deployed at CEBAF \cite{Solopova:2019}, and other laboratories \cite{Rescic:2020} to monitor trends in system behaviors precursor to faults.\\

Design of beam loss monitor networks using Correlation and Principal Component Analysis (PCA) \cite{Liu:2015} is used to determine optimum locations to place beam loss diagnostics to monitor for all known loss mechanisms in specific beamlines. Unsupervised learning techniques are used to detect faulty beam position monitors that determine beam trajectories \cite{Fol:2019}.\\

Beyond effects that directly influence beam delivery to experiments, ML techniques are being considered to assist in other critical operational aspects. Predictive schemes for equipment maintenance can be used to proactively identify components requiring attention prior to critical need. Cryogenic production and distribution will benefit from online monitoring and predictive capabilities provided by supervised and unsupervised learning by quickly detecting unplanned helium losses and alerting operators.\\

The current efforts leverage existing ML frameworks and tools. However, a detailed integration for verification, validation, and reproducibility have not been developed. Additionally, there are no current efforts to integrate uncertainty quantification into the machine learning pipeline. Finally, implementing domain aware ML, when appropriate, could provide better forward  prediction models for failure and anomaly detection. These components will be critical to provide a a full featured and reliable monitoring and prognostication system.\\ 

As more sub-systems are integrated into a comprehensive monitoring/logging framework, managing the data-load will become increasingly important. These large-scale online data sets faces a range of challenges, including multi-modal and multi-frequency high-dimensional, noisy, and uncertain input data. 
\subsubsection{Dynamic optimization of real time operation controls}
\begin{figure}
    \includegraphics[scale=0.45]{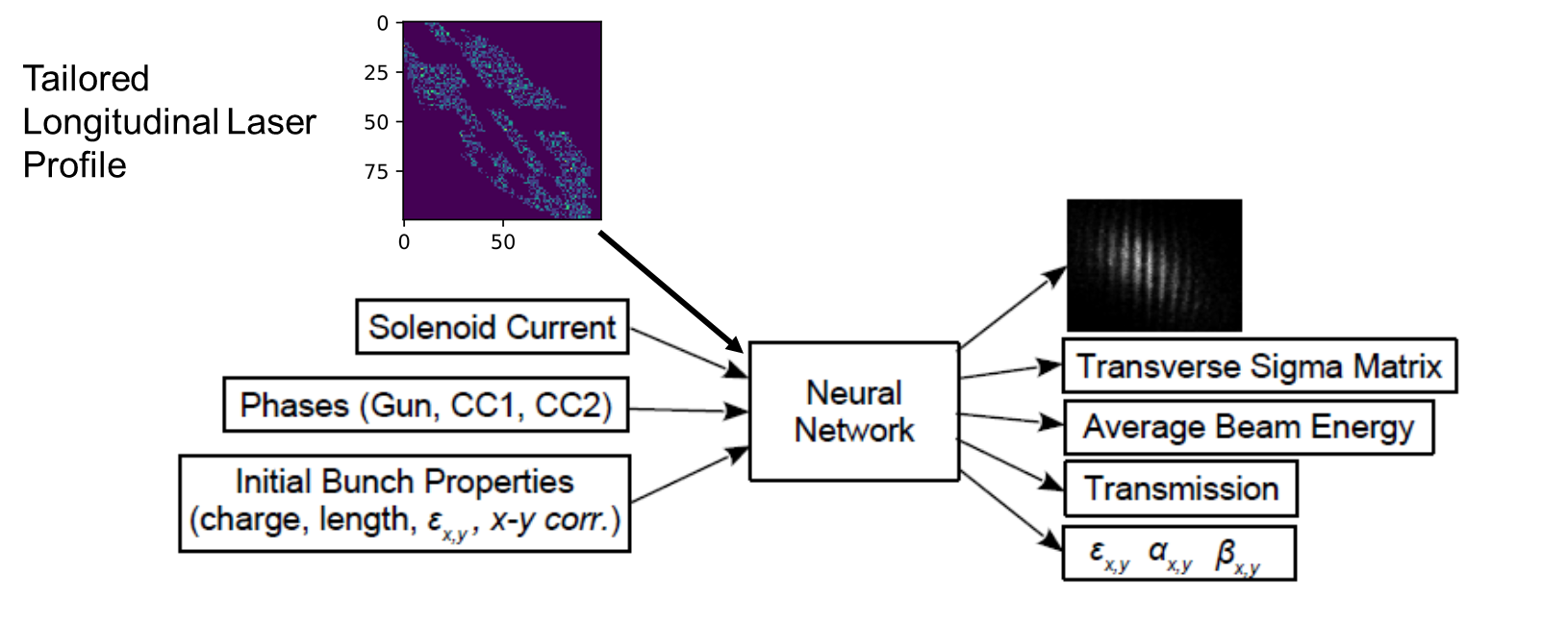},
    \caption{Neural network model used to train tunable laser profiles at Argonne National Laboratory (courtesy R. Roussel, ANL).}
    \label{fig:AWA_laser}    
\end{figure}

Frontier accelerator facilities such as FRIB and EIC will require years of operational experience to fully develop functional capabilities at their design level. AI/ML techniques are in use to improve the control over particle beams, incorporating Reinforcement Learning (RL) techniques within the accelerator control system [Schram FNAL]. Particle Swarm techniques have been tested to optimize the tuning of aperiodic ion transport lines, and are in development for advanced particle separators \cite{Amthor:2018}.
Bayesian Gaussian Processes (GP) and Neural Network (NN) methods are in use to train laser-driven photoinjector facilities in one or more degrees of freedom (Figure~\ref{fig:AWA_laser}).\\

AI/ML activities are being pursued at many NP accelerator facilities and at associated universities. These activities are mainly oriented towards addressing local issues, and are performed by individual  scientists or small teams with or without direct support from data science experts. Strategic development and deployment of AI/ML techniques across the DOE complex has high leverage of performance for investment.\\

Similar to the design optimization effort, techniques such as reinforcement learning can be use to explore the large control parameter space to dynamically optimize for real time system. Leveraging existing frameworks, such as CANDLE and ExaRL, to optimize the learning will be important, however, additional safeguards will be required to ensure that the policy network model doesn't diverge while in a real time system. The ability to process the data in a timely manner will be critical to the applicability of these techniques. Leverage and making advancement in cutting edge technology will provide the ability to deploy better models in  real time systems.

\subsubsection{Summary and Final Thoughts}
We identify specific areas of accelerator design and operations that would benefit from investments in AI and ML technologies.
\begin{itemize}

\item{\bf Data capture and streaming}
Developing a comprehensive data capturing and streaming framework will be critical to maximize the utility of the AI/ML tools. Having enough time series data from relevant sensors will be be required to build causal models that properly account for system lags, etc. As we gain confidence and understanding in these AI/ML models, moving them  closer to the sensor will allow facilities to automate parts of the operations yielding reduced downtime and operational cost. Development in AI/ML at the edge (FPGAs, etc) and model robustness will be vital. As the NP community expands its use in AI/ML  will require access to greater resources to train AI/ML models. Data aggregation and distribution to these compute resources will be an important factor.

\item{\bf Uncertainty quantification and robustness}
The need to associate uncertainties with the AI/ML predictions is critical for all efforts. However, it's particularly important when applied to Scientific User Facilities. AI/ML applications for anomaly detection and fault prediction require a quantifiable estimation of uncertainty to determine the proper coarse of action and trade-off (false positives, etc.).

\item{\bf Optimized design of accelerator systems}\\
Development and validation of virtual diagnostics (eg. longitudinal phase space monitors or predictors).  
Design and simulation of novel accelerators, and advanced engineered materials. 
Optimized diagnostic design and deployment and
improvement to beam sources and injector performance.
    
\item{\bf Improving facility performance and user experience}\\
Data-driven beam generation, transport, delivery optimization.
Automated learning for operator support.
Hardware acceleration of ML in distributed control systems.
Anomaly detection and classification and mitigation (eg. LLRF, beam diagnostics);
System health monitoring (eg. targets, cryoplant); 
Data driven system maintenance.

\item{\bf Benchmark techniques on standard models; dedicated accelerator studies}
Dedicated studies on machines and diagnostic support.
Identify specific beamlines, injectors and accelerator facilities to facilitate design and implementation of technologies, algorithms, data pipeline structures. 
\item{\bf Develop capability in AI/ML for computing at the edge (FPGA, etc.)  }
Moving AI/ML workflows closer to the sensor will allow for computing resources to be leveraged and distributed where necessary, allowing for high density data transfers to be conducted locally with reduced load on facility networks. 

\item{\bf AI cookbook of techniques and Data Science workshops/training}
Development of a community standardized toolkit for training AI/ML  scientists and provide answers to commonly encountered issues.
 
\end{itemize}

\subsection{Experimental Methods}

\subsubsection{Current Status}\label{sec:Exp:current}
AI applications to experimental applications are being developed across the subfields of NP.
In some experiments which  like those depending on image analysis, AI techniques have been 
successfully applied. This  includes the time projection chamber experiments and neutrino experiments \cite{cnn-neutrino, kuchera2019nima, Delaquis_2018, Adams:2018bvi}. Work has also been done to analyze jet substructure \cite{Lai:2018ixk}, and in detector rejection methods \cite{Barbosa:2019hux}. Current efforts expand upon this work, building on existing AI technologies. 

\begin{figure}[h]
\centering
\begin{minipage}{.45\textwidth}
\centering
\includegraphics[width=0.8\linewidth]{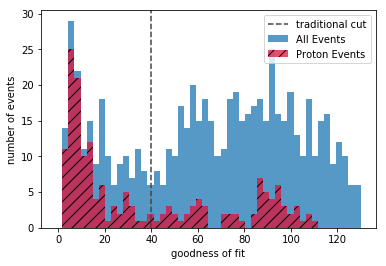}
\caption{Histogram visualizing the classification of events from the $^{46}$Ar$(p,p')$ experiments in the Active-Target Time Projection Chamber at the Facility for Rare Isotope Beams. Based on the ``goodness of fit" ($\chi^2$) distribution of the entire dataset, a cut (dashed line) was chosen at 40 (in arbitrary units). The events that were hand-classified as protons from this run are hatched \cite{kuchera2019nima}.}
\label{fig:ATTPC_pic}
\end{minipage}%
  \hfill
\begin{minipage}{.4\textwidth}
\centering
\includegraphics[width=0.9\linewidth]{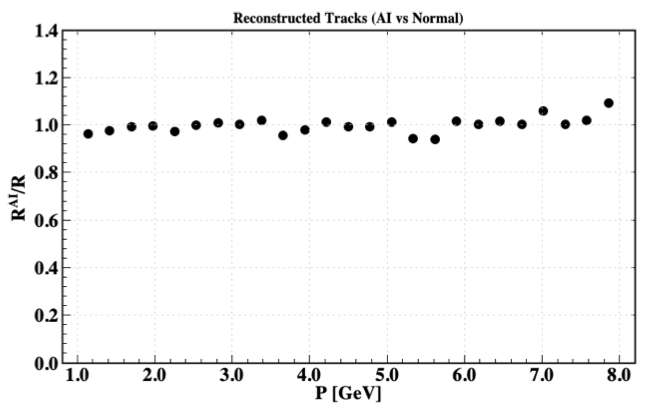}
\caption{Ratio of result from the traditional analysis method to results from a deep, fully-connected neural network for tracks in the Hall B drift chamber at Jefferson Lab. The neural network algorithm performs comparably to traditional methods, but with 6x speedup, allowing for faster analysis.}
\label{fig:HallB_pic}
\end{minipage}
\end{figure}

Significant AI endeavors in experimental NP have been in tracking in various detector setups, as highlighted above. Two examples are track classification in the Active-Target Time Projection Chamber at the FRIB and track selection in the CLAS12 drift chambers at Jefferson Lab. Figures \ref{fig:ATTPC_pic} and \ref{fig:HallB_pic} demonstrate two benefits that AI leverages over traditional methods. In the first, classification machine learning  methods were used to improve data selection over traditional cut methods. In the second, equivalent accuracy was achieved with AI methods, but with significant (6x) speedup over traditional fitting methods.






\subsubsection{Case Studies and Future Prospects}

{\bf Holistic approach to experimentation}
As a long-term, ``moonshot" vision, disparate data sources would be intelligently combined and interpreted to improve experiments. Data sources include accelerator parameters, experimental controls, and the detector data itself. Real time analysis and feedback enables the quick diagnostics and optimization of experimental setups. Accelerator-based, quick-turnaround experiments are a unique challenge in NP. ML expert systems can increase the scientific output from the beamtime allocated to each experiment. Ideally, this holistic approach can be applied to the design of the experiment itself by optimizing machine and detector properties as a single system.

{\bf Experiment design not limited by computation}
Future experimental advances in accelerator-based NP research hinges on increased luminosity, which provides the statistics necessary to observe rare processes. ML methods will reduce computational barriers to this goal. Intelligent decisions about data storage is required to ensure the relevant physics is captured. Depending on the experiment, AI can improve the data taken through data compactification, sophisticated triggers (both software and hardware-based), and fast-online analysis.

An example would be the incorporation of neural networks in the FPGAs which comprise the front-end triggers of complex experiments. The very large channel counts afforded by modern semiconductor detectors combined with high beam luminosity yield data rates that can be prohibitively demanding. Incorporating intelligent triggers with very low latency early in the signal processing chain makes this data challenge more manageable. Furthermore such triggers could act as classifiers allowing for anomaly detection on the data stream prior to the trigger decision flagging interesting events that would normally be silently discarded.  

{\bf Improving analysis}
As seen in Sec.~\ref{sec:Exp:current}, improving data analysis using ML techniques is currently proceeding with two general aims:
\begin{itemize} 
\item to use these new techniques to improve the sensitivity of current instruments and accuracy of the data, and
\item to decrease the time such analysis takes, allowing for faster turnaround time to produce scientific results.
\end{itemize}
Improving sensitivity creates more accurate datasets, which decreases uncertainty in results and increases the potential for discovery. Decreasing analysis time saves costs and allows for a higher volume of scientific output.

{\bf Uncertainty quantification}
A near term goal is to  apply AI methods with well-understood uncertainty quantification, both systematic and statistical, to experimental methods. The dominant ML algorithms used in experimental HEP and NP do not provide error estimations with model predictions, which are essential to understand experimental results. In addition,
an evaluation of metrics for the evaluation and comparison of uncertainty predictions from different models is required for widespread use of AI in experimental NP.

\subsubsection{Enabling Discoveries/What is Needed}

{\bf Educate and build a broader community }
To achieve the experimental goals outlined by the community, we must build a community of researchers knowledgeable in AI technologies. This would be greatly facilitated by centrally located, NP-supported and maintained educational resources and tutorials. Centralized resources allows for: a common foundation to build our technologies, quick dissemination of new techniques, and a bridge from available AI resources to NP related applications. 

Build an infrastructure for AI / ML scientists in the NP community. This includes laboratory positions, the establishment of university collaborations, and joint positions. 

{\bf Standardized data formats}
In order to collaborate and use AI tools effectively it is important to standardize the way we present data to these systems. Most AI tools in current use are created by industry or large open source projects with established communities. Taking on common data formats and workflows allows us to move with these communities (and each other) more quickly and effectively.


\subsection{Event Generation and Simulation}

\subsubsection{Current Status}

Simulations of physics processes and detector response are required in NP to design experiments, develop and verify analyses, and compare to theory. They are also used in theory and phenomenology to simulate data and investigate theory advances. High-precision measurements at CEBAF, RHIC, the upcoming EIC and other NP facilities require simulations with high-precision and high accuracy. Achieving the statistical accuracy needed is often computationally intensive with the simulation of the shower evolution in calorimeters being a prime example. As alternative, fast simulations with parameterizations of detector response or other computationally efficient approximations are pursued. However, they still lack the accuracy required for high-precision measurements. Here, AI provides a promising alternative. Fast generative models, e.g., GANs or VAEs, are being utilized to model physics processes and detector responses accurately and accelerate simulations. Beyond that, Bayesian optimization is applied for tuning simulations and detector design, with AI-optimized detector design being emerging for the EIC. 

\subsubsection{Case Studies and Future Prospects} 

\textbf{Accelerate simulations} High-Energy Physics has used AI, in particular GAN-based architectures, to successfully accelerate detector simulations. In some cases, in particular in case of calorimeters, the models can be directly applied to fast simulations in NP. In many cases, e.g., for particle identification detectors, new approaches to fast particle identification can be developed as, e.g., shown for Cherenkov detectors~\cite{Fanelli:2019qaq}. The resulting fast turnaround time for simulations with high-precision and high-accuracy will allow for rapid improvements of the physics reach and detection capabilities of NP experiments. 

\textbf{HPC utilization} NP experiments have few payloads appropriate to the Leadership Computing Facilities, in particular for the upcoming exascale supercomputers where accelerator technologies are being applied extensively. AI is the best near-term prospect for using accelerated hardware efficiently. Physics and detector simulations based on AI would be an ideal payload for the Exascale Computing Project. 

\textbf{AI-driven detector design} Advanced detector design requires performing computationally intensive simulations as part of the detector-design optimization process. Nowadays there are various AI-based global optimization procedures, e.g., reinforcement learning or evolutionary algorithm. Among these, Bayesian Optimization has gained popularity for its ability of performing global optimization of black-box functions which additionally can be noisy and non-differentiable. For example, an automated, highly-parallelized, and self-consistent framework based on Bayesian Optimization has been recently developed~\cite{Cisbani:2019xta}, where a PID detector for the future EIC has been considered as a case study. These studies showed an improvement in performance and provided useful hints on the relevance of different features of the detector. The same procedure can be applied to any other detector, or even combination of detectors. Also, costs can be added as parameter in the detector-design optimization process. 

\begin{figure}[t]
\includegraphics[scale=0.3]{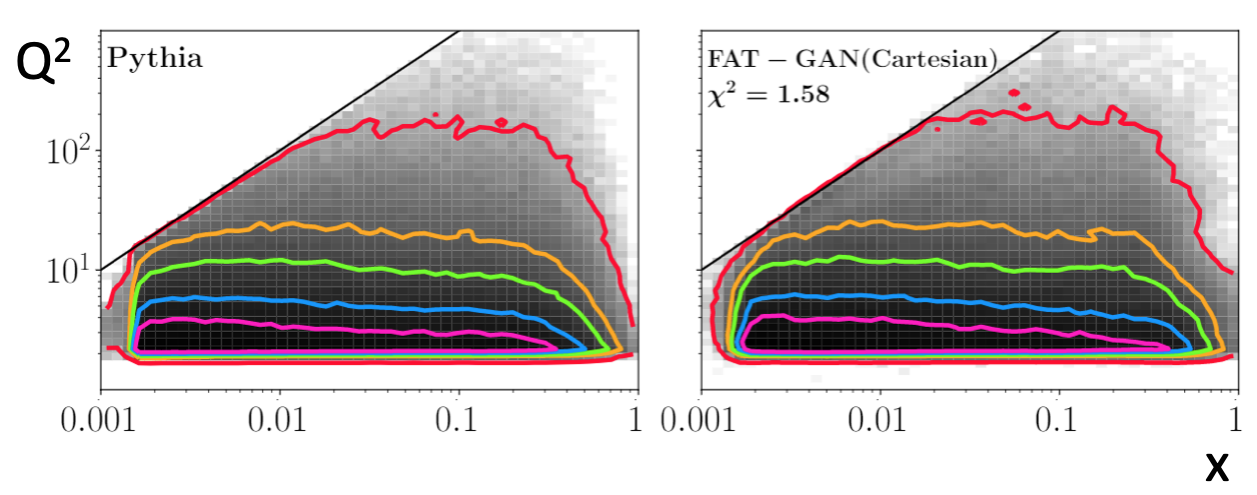}
\caption{In the pioneering ETHER project to construct Monte Carlo event generators agnostic of theoretical assumptions and phenomenological models, GANs are being developed as a repository for the behavior of the theory expressed in Pythia. As shown in the comparison of inclusive DIS kinematics from Pythia (left panel) and ETHER (right panel), the AI approach is being able to learn the Pythia data.}\label{fig:ETHER}
\end{figure}

\textbf{AI for event generators} Monte Carlo event generators describe collision processes through a combination of theory and phenomenological models. AI approaches can be applied to experimental data and map out the underlying probability distributions governing the spectrum of final-state particles in a given process. This information can be used to construct event generators in a model-independent way, providing unique ways to quantitatively test the validity of theoretical assumptions or models. Such an event generator would store the same information as that contained in the experimental data and can be viewed as compact data storage utility. A prototype event generator is currently being developed with the ETHER (Empricailly Trained Hadronic Event Regenerator) project, as illustrated in Fig.~\ref{fig:ETHER} for a comparison of Pythia generated electron-proton scattering events with those produced by a Feature-Augmented and Transformed (FAT) GAN~\cite{Alanazi:2020klf}.

\subsubsection{Enabling Discoveries/What is Needed}

AI research is multidisciplinary. An interplay of applied mathematics, computer science, and NP will facilitate the development of AI approaches to the unique questions of NP. This will allow, e.g., to design activation functions particular to NP applications or to build efficient neural networks no more complex than necessary. The multidisciplinary approach will also be helpful to understand the requirements for explainable AI and uncertainty quantification for NP simulations. To cultivate multidisciplinary AI development, access to reference data sets, as well as supplementary information for non-experts on what the NP data entails is essential.

\subsection{Bayesian Inference for Quantum Correlation Functions}

\begin{figure}[h]
\includegraphics[scale=0.5]{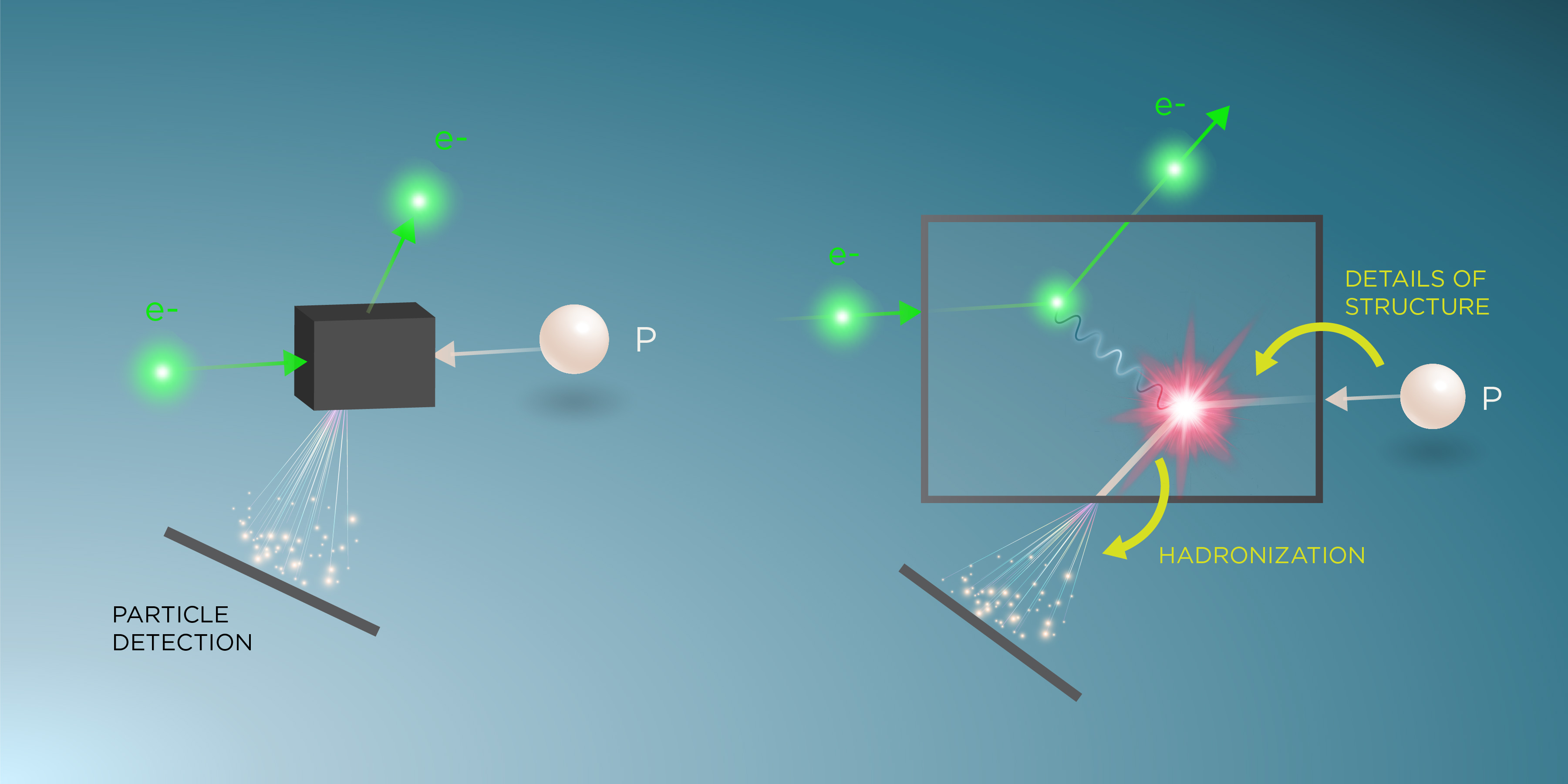}
\caption{Illustration of factorization in electron--proton scattering. The black box is interpreted on the right in terms of short-distance reactions of quarks and gluons, with the details of proton structure and hadronization parametrized in terms of QCFs.}
\label{fig:QCF-factorize}
\end{figure}

Determining the 3-dimensional ``tomographic'' structure of the proton and nuclei in terms of the elementary quark and gluon (or parton) degrees of freedom of QCD remains one of the central challenges in modern NP.
A fundamental complication in this endeavor is the fact that quarks and gluons always remain confined inside hadrons and never observed directly in experiments.
This constitutes a classic ``inverse problem'': how to reliably infer the quantum correlation functions (QCFs) that characterize hadron structure and the emergence of hadrons in terms of partons from the experimental data --- Fig.~\ref{fig:QCF-factorize}.

Existing approaches to extract QCFs, such parton distribution functions (PDFs), fragmentation functions (FFs), transverse momentum dependent distributions (TMDs) or generalized parton distributions (GPDs), from data rely on Bayesian likelihood inference, coupled with suitable parametrizations of the distribution functions on the internal parton momenta.
The complexity of mapping between the large quantities of high-precision data expected from JLab~12~GeV (as well as from the future EIC) and the multidimensional QCFs, many of which have never been been explored, will require the creation of a new paradigm in order to assess the impact of the data.
An important opportunity therefore exists for utilizing AI/ML techniques to develop the next generation of QCD analysis tools that can more efficiently map between observables and QCFs and maximum the science output from future facilities.

\subsubsection{Current Status}

Historically the extraction of 1-dimensional QCFs, such as PDFs or FFs, has relied on the maximum likelihood method, which is adequate for cases involving a small number of distributions, but can introduce significant bias and error when applied to more complicated problems involving multidimensional functions.
Current state-of-the-art analyses seek to overcome these problems by employing Monte Carlo sampling (NNPDF \cite{AbdulKhalek:2019ihb} and JAM~\cite{Sato:2019yez} Collaborations) to take into account the multiple solutions, and {\bf simultaneously} determining various types of QCFs which appear in different observables to account for feedback effects~\cite{Ethier:2017zbq}.

Other examples of state-of-the-art techniques currently employed for 1-D QCF studies include the use of {\bf neural net} methodology for proton PDFs \cite{AbdulKhalek:2019ihb}, and the application of generative adversarial networks (GANs) for mapping PDFs \cite{Rojo:2018qdd}.
In the transverse momentum sector, the first global TMD analysis was performed recently \cite{Cammarota:2020qcw} using the JAM MC methodology extended to the 3-D sector.
Exploratory studies of fitting GPDs with neural networks were made for a limited set of deeply-virtual Compton scattering data~\cite{Kumericki:2011rz}, and recently the more general approach of parametrizing Compton form factors (integrals of GPDs) with neutral nets has been explored~\cite{Kumericki:2016ehc}.
Finally, as lattice QCD simulations at physical quark parameters are becoming more feasible, synergies between global QCD analysis of experimental data and lattice results are being actively explored~\cite{Lin:2017snn}, including the first attempts to perform simultaneous fits to measured cross sections and lattice matrix elements of nonlocal operators, whose Fourier transforms are related to PDFs.

\subsubsection{Case Studies and Future Prospects}

The history of applying ML tools to study the hadron substructure is rather brief.
A recent example used neural nets to construct a {\bf universal Monte Carlo event generator} (UMCEG) for electron-proton scattering, that is free of theoretical assumptions about underlying particle dynamics \cite{Alanazi:2020klf}. 
This project, funded by the Jefferson Lab LDRD program, applied generative adversarial network (GAN) technology to simulate particle production at the event level.
A new {\bf feature-augmented and transformed GAN} (FAT-GAN) was developed to select a set of transformed features from particle momenta (generated directly by the generator), and use these to produce a set of augmented features that improve the sensitivity of the discriminator. 
The new FAT-GAN was tested on pseudodata generated by the Pythia event generator \cite{Sjostrand:2007gs}, and was able to faithfully reproduce the distribution of final state electron momenta in inclusive electron scattering.
The FAT-GAN strategy can be generalized to GANs for simulating other reactions under different conditions, as well as learning exclusive events, and alternative strategies, for example using convolutional neural networks (CNNs), can also be explored.

\begin{figure}[h]
\includegraphics[scale=0.5]{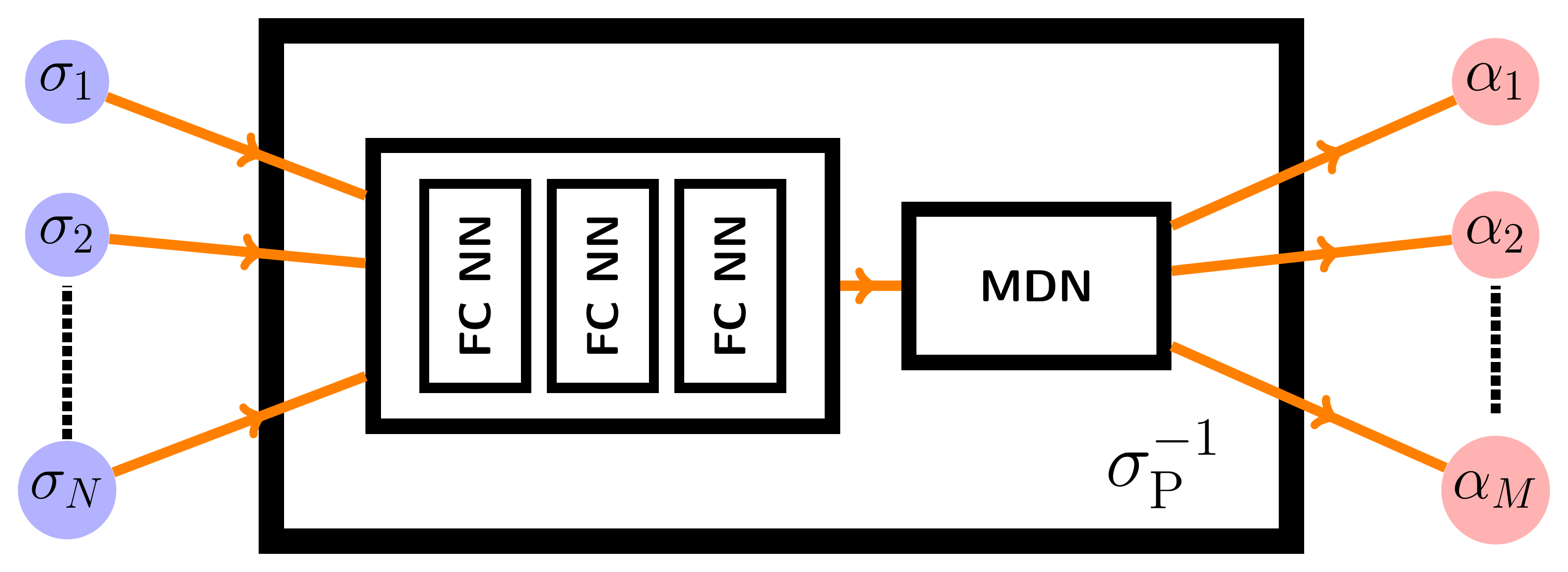}
\caption{AI architecture for inverse mapping connecting an $N$-dimensional space of observables $\sigma_i$ into an $M$-dimensional space of parameters $\alpha_j$. Aside from the fully connected neural networks (FC NN), the inverse mapper is equipped with a mixture density network (MDN) layer to allow for possible multiple solutions for the inverse problem.
The subscript ``P'' differentiates the parametrized inverse function $\sigma^{-1}$ from the true inverse function.}
\label{fig:QCF-inverse-map}
\end{figure}

Another important recent application of AI has been to the development of inverse mapping methodology using machine learning for Bayesian inference of QCFs --- see Fig.~\ref{fig:QCF-inverse-map}.
Two machine learning prototypes have been explored, based on a mixture density network and a parameter-supervised autoencoder, which have been tested and validated first on a toy model for inclusive DIS, and subsequently on a real global analysis of DIS data. 
The prototypes were found to be capable of mapping PDFs to within 1-$\sigma$ CL, consistent with those found in recent global Monte Carlo fits~\cite{Sato:2019yez}.
Extension of the methodology to the 3-D sector remains an important future challenge.

\subsubsection{Enabling Discoveries/What is Needed}

To maximize the potential benefit from AI for QCF inference studies, collaboration between QCD physicists and machine learning experts is needed in order to translate the domain knowledge of QCD into generic problem definitions that can be addressed with cutting-edge AI technology.
To this end, the creation of {\bf joint positions} between NP and AI will promote cross-disciplinary fluency in both fields.

The development of an interactive {\bf web-based global analysis} platform to perform global QCD analysis ``on the fly'' will allow users to study how different setups (choice of specific data sets or kinematic regions, or improvements on data uncertainties from future facilities, such as the EIC) can affect the inferred QCFs.
The vision is to move from the limited paradigm where QCFs are numerically tabulated at interpolation grids, with rigid connections between the data and QCFs, to a more flexible paradigm where QCFs can be generated dynamically from user input.

The creation of such web-based analysis infrastructure would be a valuable tool for the NP community, but will require identifying the most efficient computing platform to host such a service and computing resources for its realization.
There is also a critical need for {\bf production-level hardware resources} to enable the analysis of the large quantities of high-precision data expected from new experimental facilities, in order to understand the deep connections between the data and the QCFs.

\subsection{Additional Contributions Received}
\subsubsection{Relativistic Heavy Ions}

At extremely high temperature or density, quarks and gluons become deconfined and form a new state of matter -- Quark Gluon Plasma (QGP). One can study this matter through high energy nuclear collisions at Relativistic heavy ion collider (RHIC),  Large hadron collider (LHC) and other facilities, as well as computer simulations by analyzing the four-momenta and species of final state particles produced in each single collision. The dynamical evolution of the collision systems can be described by hybrid models with relativistic hydrodynamics and hadronic cascades at different stages of the collision. One can infer the initial state of the collision and the intermediate evolution from comparisons the data on final state particles from experiments and simulations.  

AI plays an important role in compressing the high dimensional heavy-ion collision data to low dimensions, extracting the model parameters and their uncertainties with Bayesian analysis, classifying the equation state, regressing the initial nuclear structure or in solving partial differential equations of relativistic hydrodynamics using deep neural networks. These AI applications are described in the following.

\begin{description}

\item[Compressing data to low dimensions] Many experimental observables are designed to compress complex high energy nuclear data to low dimensions
using simple projection, statistical mean, variance and correlations along a few directions.
Unsupervised learning algorithms such as PCA is widely used in the field of high energy nuclear physics,
to automatically extract the most informative features in data.
PCA can be used to determine the magnitude of different longitudinal fluctuation modes \cite{Bhalerao:2014mua},
which helps to constrain the initial state entropy deposition along the beam direction in heavy-ion collisions.
Since the initial state fluctuations of entropy density in the transverse plane is converted to final state correlations of particles in momentum space,
 the collectivity and anisotropy of final state particles along the azimuthal angle direction are quantified by the flow harmonics $v_n$.
The $v_3$ factorization breaking is well described using 2 initial state fluctuation modes given by PCA and a linear hydrodynamic response \cite{Mazeliauskas:2015vea}. PCA also rediscovers flow harmonics \cite{Liu:2019jxg} which are originally computed from Fourier decomposition.

\item[Bayesian analysis to extract QGP properties]
Bayesian analysis uses the likelihood between low dimensional experimental data and model output to constrain model parameters, such as the QCD equation of state \cite{Pratt:2015zsa}. The prior QCD EoS used in hydrodynamics is parameterized to cover the physical equation of state functional space.The posterior distribution of the EoS agree with lattice QCD calculations.
To take into account the effect of other entangled parameters,   Trento + iEBE-VISHNU + UrQMD model is used to do a global fitting using Bayesian analysis \cite{Bernhard:2016tnd,Bernhard:2019bmu,Paquet:2020rxl}. The clear peak structure in the posterior distributions of model parameters indicates non-zero shear and bulk viscosity of the QGP.
When high energy partons traverse through QGP, they loss energy by elastic scattering and gluon radiations.
The Bayesian analysis is also used to constrain the heavy quark diffusion coefficients \cite{Xu:2017obm}, the light quark $\hat{q}$ \cite{Soltz:2019aea} and the jet energy loss distribution \cite{He:2018gks}.

\item[Jet classification in heavy ion collisions] 
The applications of neural network was used in 1996 to determine the impact parameter of heavy-ion collisions \cite{Bass:1996ez}, with  a one-hidden layer neural network. Various architectures of deep neural network are used in jet flavor classification for proton-proton collisions. However, the applications to heavy-ion jet classification is rare. The classification performance worsens due to soft gluon radiations affecting soft jet substructure \cite{Chien2019}. Recently a point-cloud-like network called particle/energy flow network is employed in jet flavor classification \cite{Komiske2018} and is used to design new physical observables for heavy-ion jets \cite{Lai:2018ixk}.

\item[Classification for nuclear phase transition]
Beam energy scan (BES) project aims to locate the QCD critical point that separates the first order phase transition and smooth crossover in the QCD phase diagram by colliding heavy ions at various energies. Deep convolution neural network is used to classify these two different nuclear phase transition regions \cite{Pang:2016vdc} using relativistic hydrodynamic simulations of heavy ion collisions. The phase transition type used in the equation of state is encoded in the evolution and deep neural network helps to decode this information from the complex final state output of heavy-ion collisions. Although there is entropy production and information loss, the network succeeds in classifying nuclear phase transition types with approximately 93\% accuracy. Deep convolution neural network uses images as input, a more natural representation of the heavy-ion data of a list of particles with their four momenta, pid and charge information. Point cloud network is a perfect architecture for this data structure. A recent study uses point cloud network to classify Spinodal and Maxwell constructions of the first order phase transition \cite{Steinheimer:2019iso}.

\item[Regression for nuclear-shape deformation]
 Most heavy ions used at RHIC and LHC are deformed. The collisions of deformed nuclei produce complex correlations between charged multiplicity and anisotropic flow. Using Monte Carlo simulation data, a 34-layer residual network is used to predict the values of nuclear shape deformations \cite{Pang:2019aqb}. The network succeeds in predicting the magnitude of nuclear shape deformations but not their signs, which indicates that there is a degeneracy between high-energy collisions of prolate-prolate and oblate-oblate nuclei.

\item[Interpretation and explanation] 
Interpretation is important in understanding what has been learned by the black box deep neural network. In the classification task for nuclear phase transition, a prediction difference analysis algorithm is used to locate the most important phase space regions in the input for classification. In the regression task for nuclear shape deformation, a regression attention mask algorithm is developed to highlight the regions that are important for the decision making.

\item[Accelerate relativistic hydrodynamic simulations]
Accumulating data in heavy-ion collisions is slow. Stacked-UNet is used to solve relativistic hydrodynamic equations \cite{Huang:2018fzn}. The time evolution of the energy density and fluid velocity from neural network method agree with 2+1D viscous hydrodynamics. The trained network can solve hydrodynamic equations 600 times faster than numerically solving partial differential equations. As a comparison, the GPU parallelization brings 60 to 100 times speed up.

\end{description}

Current study of heavy-ion collisions with machine learning have used data set generated with model simulations. To apply these techniques to real experimental data, one has to taken into account the acceptance and efficiencies of the detectors. This can be accomplished through incorporation of the characteristics of the detector in the model simulations which are used to train the network for final application to real experimental data. In the meantime, advance in the accelerated model simulations with more realistic physics scenarios are needed for more robust AI studies. 

\subsubsection{Project~8}
\begin{figure}[h!]
\centering
\includegraphics[width=\textwidth]{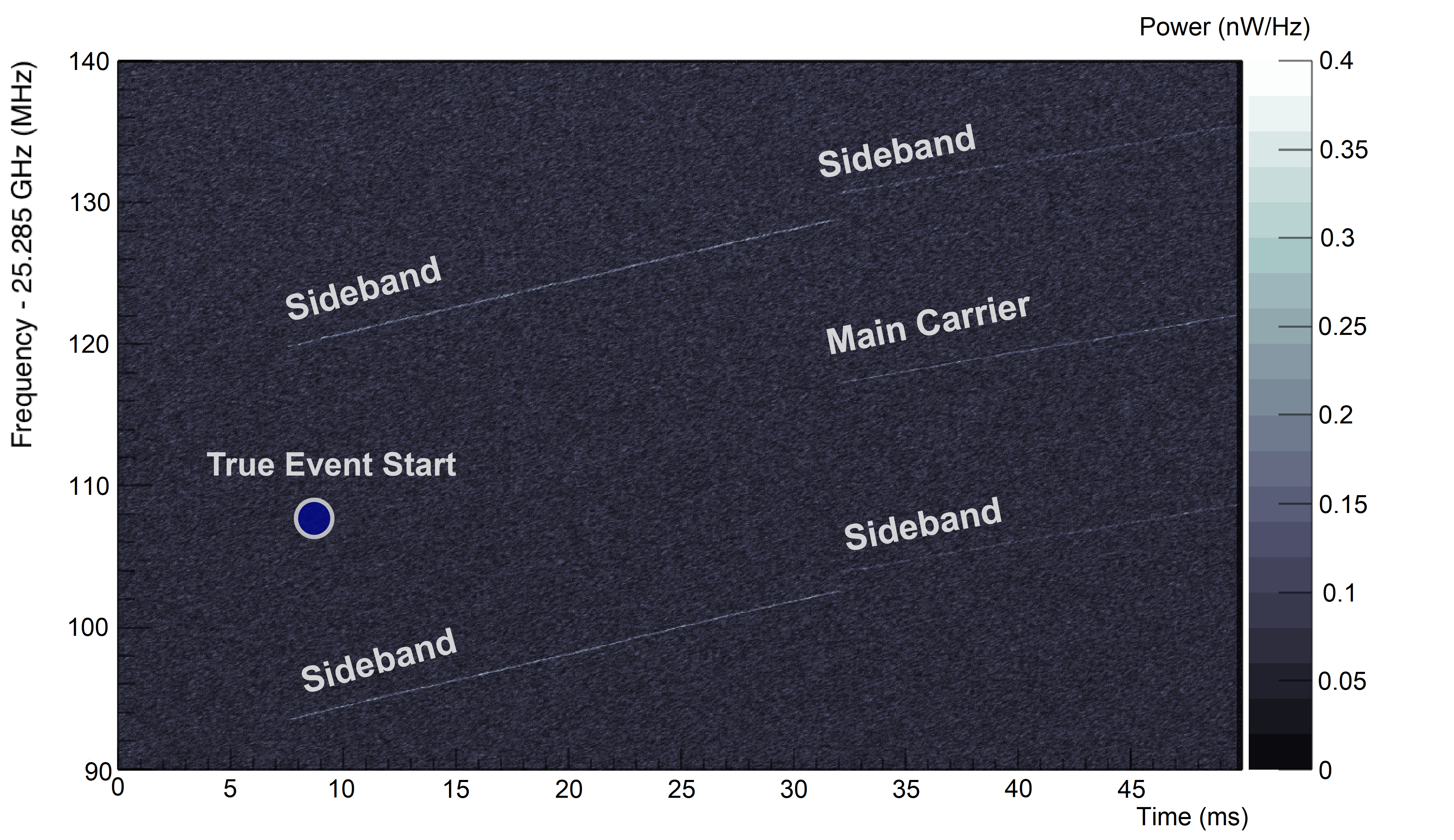}
\caption{A multi-track CRES event featuring five visible tracks.  In the first set of tracks, on the left, the main carrier track is not visible; it is important to know that the visible tracks are sidebands so that the initial track start can be determined accurately.  Figure is from~\cite{ref:p8_signal_classification}.}
\label{fig:tracks}
\end{figure}

The Project~8 collaboration is developing an experiment to measure the absolute neutrino mass with cyclotron radiation emission spectroscopy (CRES).  The event reconstruction process for Project~8 can be framed as a challenge of feature recognition in noisy data, where the features to find are the electron tracks and how they are grouped together.  The Project~8 collaboration has studied two uses of machine learning to improve track and event reconstruction. The first application was to differentiate different types of tracks by their characteristics~\cite{ref:p8_signal_classification}.  Figure~\ref{fig:tracks} shows an electron event with five visible tracks.  The four sideband tracks and one visible main-carrier track are labeled.  We first analyzed individual tracks and extract parameters like slope and power density, and then applied a Support Vector Machine to distinguish three track populations: main carrier tracks with high pitch angles (the angle of the electron's momentum relative to the magnetic field in the experiment), main carrier tracks with low pitch angles, and sidebands.  Having this information can help in reconstructing events, avoiding problems that might occur when particular tracks are not observed, like the missing main carrier in Figure~\ref{fig:tracks}.

Machine learning can be applied to Project~8 data to identify tracks, as well.  We are developing a method for identifying tracks using a Convolutional Neural Network (CNN).  This particular task is a straightforward application of a CNN with a U-Net architecture.  Such a tool, once optimized, will be used to do the initial optimization of the tracks in events such as Figure~\ref{fig:tracks}.  While the initial application of the CNN to Project~8 data is straightforward, there are a variety of details to establish, such as accounting for all of the necessary track topologies, and understanding the efficiency of detection.

\subsubsection{NEXT}
The NEXT neutrinoless double beta decay program has as its primary physics goal discovering or severely limiting parameter space for the Majorana nature of neutrinos in 136Xe decays. NEXT will undertake this search in a staged program of high pressure gas xenon Time Projection chambers (HPgXeTPCs), culminating in a multi-ton detector that will be effectively background-free. 

HPgXeTPCs, because of the benefits of gaseous xenon (perhaps with a He additive) including its small Fano factor, allow to see the topology of the double beta decay while achieving sub 1$\%$ energy resolution. The technology and the path to the necessary low background model has been demonstrated in a small detector NEXT-NEW. A future design of the 1-ton-scale High Definition (HD) design is shown below, along with a typical double beta event in simulation with its Bragg peaks at the end of each track. The exquisite topological information in these detectors calls out for Deep Convolutional Neural Nets (DNNs) to perform tasks such as signal and background classification and, in fact, full semantic segmentation-based event reconstruction.

The collaboration has already published \cite{Renner_2017} work on DNNs applied to NEXT-NEW data. A team is now at work on its Summit  allocation to extract optimal sensitivity from simulated ton-scale designs. The team has already shown  its effective use of sparse DNNs in similar highly-parallel applications on Summit, and early work is already bringing benefits to the extremely promising (multi) ton NEXT program.

\subsubsection{WANDA}

The sequence of steps whereby nuclear data is compiled, evaluated, processed, and incorporated into applications is referred to as the ``Nuclear Data Pipeline”''. The pipeline provides the critical connection between laboratory measurements and their eventual use in models of reactors, isotope production, detectors for non-proliferation, supernova explosions, radchem networks, and many other systems. To improve and supercharge this pipeline, the nuclear data community has extensive needs, including: more rapid, accurate, and robust evaluations; quicker compilation of data and accompanying contextual information from published experimental work; robust methods to optimize experimental design for verification, validation, and benchmarking; wider use of realistic physics models in transport simulations; and reproduction of the results of complex multi-physics codes via fast-execution surrogate models. 
AI/ML tools have tremendous potential to address all of these critical needs. During the recent Workshop for Applied Nuclear Data Activities (WANDA) \cite{WANDA}, the nuclear data community has recently identified  a number of key areas in which AI/ML advances have already made significant impacts and show substantial promise both in the short term and long into the future. Targeted investments are needed now to fully realize the potential of AI/ML in nuclear data, preferably by leveraging AI/ML advances in other areas for use in nuclear physics and simultaneously driving AI/ML innovations. Some of the many areas to emphasize include: 
\begin{itemize}
\item
Using AI/ML tools to identify systematic trends in nuclear data that were missed by human evaluators, and developing AI/ML emulators to incorporate complex physics models into evaluations, so that evaluations can be more robust, new physics can be uncovered, and predictive power can be improved. 
\item Exploiting AI/ML tools to process complex relationships between nuclear data and integral experiments to develop rigorous validation approaches, so that that AI/ML tools can be confidently deployed in nuclear energy, nuclear security, and other applications where safety is paramount.
\item Quantifying the intrinsic uncertainties of AI/ML tools, so that their results can be fully integrated into the nuclear data UQ process that is critical to the validation, verification, benchmarking, and normalization activities widely utilized across nuclear data activities.
\item Developing a new, standardized, QA-vetted, well-characterized database of nuclear information including UQ that can be easily input into AI/ML codes, so that AI/ML advances can be more quickly used across the nuclear data pipeline.
\item Collecting and sharing fitted models, training data, and notes on their applicability and limitations, so that the reproducibility of AI/ML results can be enhanced and advances from across disciplines can be best leveraged for the widest utilization.
\item Using AI/ML tools to both develop surrogate physics models and use them to sequentially search and optimize over a wide space of experimental design, so that the most impactful data are targeted and collected more efficiently, and so that specific deficiencies in data needed for robust evaluations are avoided. 
\item Developing natural language processing (NLP) tools to automatically compile new results, so that errors in data entry can be reduced, consistency checks can be facilitated, expert validation and verifications can be quickly done, and database insertion can be seamlessly performed. 
\item Fostering collaborations between nuclear researchers and AI/ML experts, so that appropriate algorithms can be efficiently determined for a given problem and subsequently trained, tuned, and deployed for maximum scientific impact while minimizing biases or unphysical results.
\end{itemize}

\section{Cross Cutting Topics}
The breakout sessions of the workshop focused on topics in NP with the knowledge that there are commonalities between these topics, and indeed, across many scientific domains. These 'cross cutting' areas span the spectrum from the development of methodologies and mathematics for AI approaches, the need for sophisticated data management and curation, and means to establish an AI cognizant workforce.  The following section outlines a number of these cross cutting topics.

\subsection{Statistical methods and tools}\label{AITools}

        Statistics and statistical methods are based on probability spaces, defined in terms of sets and probability measures. They aim to provide a better understanding and quantified characterization of a given set of data.  Data mining uses statistics as well as other methods to find patterns in order to explain phenomena.  Machine learning uses data mining and other learning algorithms in order to predict future outcomes, and AI uses models based by machine learning to make intelligent decisions.  As the application of AI to NP is in the early stages, the integration of statistical methods and uncertainty quantification into more advanced AI applications is still under development. 
        
        \subsubsection{Overview of approaches in NP}
        
        In lattice QCD and other lattice field theories, AI has been used for configuration generation, propagator inversion, observables, and overcoming the sign problem.  Among the various statistical methods utilized, Jensen-Shannon divergences have been used to distinguish gauge field ensembles using deep neural networks \cite{Shanahan:2018vcv}. 
        Bayesian neural networks have been used for spectral reconstruction \cite{Kades:2019wtd} and reconstructing parton distribution functions \cite{Karpie:2019eiq}. Machine learning regression errors for parton distribution functions has also been quantified using bias correction and bootstrap resampling \cite{Zhang:2019qiq}.  Bayesian inference and other statistical methods applied to model parameterizations of partonic structure are key to extracting parton distribution functions, fragmentation functions, transverse momentum dependent distributions, and generalized parton distributions.
        
        In low-energy nuclear theory, Bayesian methods have been used across a variety of different problems for uncertainty quantification.  This includes Bayesian calibration for nucleon-nucleon phase shifts \cite{Wesolowski:2018lzj} and direct nuclear reactions \cite{King:2019sax,Lovell:2018bao,Catacora-Rios:2019goa}; Bayesian Gaussian processes for truncation errors in effective field theory \cite{Melendez:2017phj,Melendez:2019izc} and uncertainties in neutron-alpha scattering and three-body parameters \cite{Kravvaris:2020lhp}; Bayesian calibration for $A$-body calculations \cite{Ekstrom:2019hlw} and mass models \cite{McDonnell2015,Kejzlar:2020vla}; Bayesian extrapolations \cite{Neufcourt2018}; Bayesian model averaging  \cite{Neufcourt2019a,Neufcourt2020,Neufcourt2020a}; and Bayesian neural networks for r-process beta decays \cite{Niu:2018trk}, alpha decays \cite{Rodriguez:2019rnj,BanosRodriguez:2019qgm}, and spallation cross sections \cite{Ma2020r}.  Bayesian regularization as well as other approaches have been used for uncertainty quantification in applying neural networks to applications such as the extrapolation of truncation errors in nuclear structure calculations \cite{Jiang2019,Negoita2019} and variational methods \cite{Keeble:2019bkv}. 
        
        AI applications to experimental NP are being developed across the subfields of NP. Experiments that map well to existing AI technologies, such as image analysis problems, have demonstrated success in NP. Examples include time projection chamber experiments and neutrino experiments \cite{cnn-neutrino, kuchera2019nima, Delaquis_2018, Adams:2018bvi}. Work has also been done to analyze jet substructure \cite{Lai:2018ixk}, and in detector rejection methods \cite{Barbosa:2019hux}. Current efforts are expanding upon this work, building on existing AI technologies. 
        In the future one would like to apply AI methods to experimental methods with systematic and statistical uncertainty quantification.  Similarly, Bayesian optimization will also be extremely useful for tuning event simulations and detector design.
        
        AI techniques in accelerator science and operations have been adopted and utilized for some time. Early usage of simulated annealing and genetic algorithm techniques \cite{Lidia:1995, Chubar:2007} were applied to optimize the distribution of pure permanent magnets in extended undulator assemblies for SASE FELs. SVD techniques have been used to optimize steering control in storage rings \cite{Corbett:1993}. Genetic algorithms and optimization techniques have been used to brightness of electron photoinjectors \cite{Bazarov:2005} and electron synchrotrons and storage rings \cite{Gao:2011}. Recent reviews \cite{Hofler:2013, Edelen:2019} have identified uses of artificial neural networks, convolutional neural networks, Bayesian optimization, reinforcement learning, random forest, and other methods in accelerator controls \cite{Edelen:2016}, longitudinal phase space prediction \cite{Emma:2018}, anomaly detection in SRF cavities and beam diagnostics \cite{Solopova:2019,Fol:2019}, FEL performance enhancement \cite{Scheinker:2019}, etc. Current efforts are expanding in all areas of accelerator control, optimization and design, diagnostics and prognostics.

    \subsubsection{Use of Current Tools}  
    The current surge in AI has provided great advances in software tools and hardware that can provide the basis of machine learning systems used in data processing. Readily available off the shelf solutions are well suited for basic classification problems, particularly for images. Analysis of experimental data however, requires regression networks that often need careful tuning to specific problems and data sets. In addition, scientific results require well understood systematic uncertainties in values obtained from any analysis. For example, charged particle tracking requires not only a 5 parameter state vector, but also a 15 parameter diagonal covariance matrix to represent its uncertainties. These are needed as inputs to kinematic fitting routines which combine constraints imposed by physics with the experimentally measured values in order to achieve optimal resolutions. Scientific results also require study to ensure no bias is introduced by the analysis technique. More so than is needed by industrial applications. 

\subsection{Collaborations and collaborative activities}\label{Collaborations}

The importance of community and collaboration were a cross cutting theme in the workshop.  There are distinct types of collaboration, each of which is beneficial. 
 
\subsubsection{NP Communities of Practice}

Relative to other communities, the NP scientists have relatively few communities of practice that enable knowledge exchange on technical topics or the ability to articulate the requirements for community developed and supported tools.  This is in contrast to, for example, the HEP community, which has several sanctioned or funded activities that focus on computing, including the HEP Software Foundation (HSF) (sponsored by CERN), the IRIS-HEP collaboration (funded by NSF) and the Forum for Computational Excellence (funded by OHEP).  HSF in particular was instrumental in developing computing focused white papers for the the European Strategy for Particle Physics.

The scientific expertise to approach the challenges in NP lies  within the NP community.  For that reason, communities of practice within NP to share knowledge computing knowledge could be invaluable tools towards addressing common challenges.   These are happening on a small scale, such as the Jefferson Lab AI lunch series and the Monthly Computing Round Table hosted jointly by BNL and Jefferson Lab.  Community based groups could serve as a clearing house for training opportunity announcements and similarly to the HSF, as a tool for organizing community white papers.  One of the outcomes of the AI for NP Workshop is the establishment of a proto-community that came together to produce this report.

Several concrete actions that could be undertaken by a community of practice is developing a portal for community based A.I. training resources and the development of AI recipe books.  Another topic could be a discussion around data management standards. Extending these activities to include AI experts would be beneficial to creating a much needed community to leverage the rapid advancement of methods and tools in the AI/ML communities.

\subsubsection{Engagement with Data Science community}
The NP community recognizes the importance of engaging the data science community to develop technologies that enable innovation in NP. Research groups have begun collaboration with computer scientists with demonstrated success~\cite{Alanazi:2020klf}. However, a broader effort to formally collaborate with the AI scientists can advance AI technologies in NP while taking advantage of unique aspects of data in NP to inform innovation in AI.  Fostering such collaboration is essential for long term success in developing AI techniques that realize the potential for impacting NP challenges.  

In order to best interact with the AI community, both parties must identify and engage in mutually beneficial research topics. This requires education and interaction of the two fields. To maximize collaboration, laboratories and institutions can create an infrastructure in which AI scientists are an integral part of the field.  This can be accomplished through joint projects that includes well-defined metrics of success for an AI scientist working in a physics field. The collaboration will be mutually beneficial, with the AI work not considered a service, but as a true collaboration. This can be evidenced by nuclear physicists and AI scientists publishing together, whether in physics or AI journals. Building a merged community of physicists and data scientists brings challenges in nuclear physics data analysis to the consideration of AI researchers as they develop new methods. This will allow AI technology to advance in line with our community’s needs.


\section{Engagement with ASCR}
For the past few years, the U.S. Department of Energy, Office of Science program in Advanced Scientific Computing Research (ASCR) has been conducted several workshops directly and indirectly focused on AL/ML which resulted in several reports.\\ \\
In January 2018, the ASCR Basic Research Needs workshop on Scientific Machine Learning \cite{BasicResearchNeeds} identified six priority research directions (PRDs). The first three focused on the foundation research themes: 1) Domain-awareness, 2) Interpretable, and 3) Robust. Within the NP community, the use of domain aware ML to leveraging scientific domain knowledge by enforcing physical conservation law and governing equations was identified. Additionally, providing robust ML solutions is important for scientific research and critical when deployed at scientific user facilities (SUFs).
The last three focused on capability research themes: 4) Data-Intensive, 5) Enhanced Modeling and Simulation, 6) Intelligent Automation and Decision Support. All three PRDs of these items have clear applications within the NP community. For example the semi-automation of emerging SUFs could significantly reduce operational cost and downtime.\\  \\
Although not explicitly focused on ML, ASCR convened a workshop on in situ data management (ISDM) on January 28–29, 2019 \cite{InSitu}. The goal of the ISDM workshop was to consider in situ data management to support traditional and future scientific computing needs. Six PRDs were identified: 1) Pervasive, 2) Co-designed, 3) In Situ Algorithms, 4) Controllable, 5) Composable, and 6) Transparent.
These priorities are of particular interest to DOE NP since they could directly feed into the existing and future facilities, such as FRIB and the EIC.
In particular, the need for provenance and reproducibility was explicitly linked to the development and use of ML. Additionally, in situ algorithm and controllable ISDM would enable semi-automated SUFs.\\ \\
On June 5, 2019, the DOE Office of Science (SC) organized a one day workshop centered on the topic of \emph{Data and Models: A Framework for advancing AI in science} report \cite{DataModels}. Three priority opportunities were identified: 1) democratize access to benchmark science data, 2) make AI operational in science with composable services and 3) address open questions in AI with frameworks. Providing a Findable, Accessible, Interoperable, and Reusable (FAIR) dataset and composable tools would accelerate the ability for members of the NP community to develop new algorithms and efficiently train them using these services.\\ \\
Between July and October 2019, four town hall meetings dubbed ``AI for Science'' were conducted to discuss and identify the scientific needs and opportunities across a diverse collection of domains (biology, physics, mathematics, accelerators, computing, etc.). Some of the most notable grand challenges for NP included: 
Automate and/or optimize the operation of accelerators and detector systems;
Improve experimental design and real time tuning. \\ \\
Finally, it was identified that the NP community could benefit from using existing AI/ML solutions by leveraging existing ASCR investments. For example, the Exascale Computing Project (ECP) has created tools that can accelerate computationally expensive tasks. For example, using CANDLE to perform large scalable hyper-parameter optimizing scans could potentially significantly improve on existing results. Similarly, the ECP ExaLearn project is now developing scalable tools to address common AI/ML challenges such as developing surrogate models, inverse problems, and automated design and controls challenges. These tools could save a significant amount of development time and allow the NP community to focus on solving domain specific challenges.\\ \\
As the NP community expands its use in AI/ML it will require access to greater computing resources to train AI/ML models. The NP community should leverage the existing ASCR computing facilities and develop a data aggregation and distribution community plan.

\section{The importance of Data Management}

AI techniques are reliant on the quantity and quality of the data and for this reason, applications of AI are likely to result in a paradigm shift in data management.  Accessibility of the data to the wider NP community would create a connectivity across experiments that could increase collaboration.  Viewing data as a valuable commodity impacts decisions on how data from experiments and simulations is collected, cataloged and accessed.   AI techniques could also facilitate near real-time calibration and analysis.

As mentioned in several of the summaries, current analysis techniques often 'flatten' the experimental data.  To maximize the usefulness of the data, it will be important to have agreements and documentation on 'processing' of experimental data, the application of theoretical assumptions and the treatment of systematic uncertainties that will be used as training samples or as part of combined analysis.   All relevant information about the data will have to be stored with the data.   This should trend towards the development of appropriate standards consistent with FAIR data principles and frameworks that capture data and metadata. 



\section{Workforce development}\label{Workforce}

\subsection{Education}

There are only 26,000 AI researchers currently in the US. This is estimated to represent only a fifth of the current demand. There is an urgent need for training in AI, at a variety of educational levels and for diverse audiences. To this end,there is an urgent need to develop a range of outreach, recruitment, and educational activities. 
NP research will serve to raise interest in AI-related fields. The goal is to retain talented students in AI-related fields and to help them to secure employment in a wide range of careers, thus ensuring that the new techniques and concepts developed in NP laboratories are widely disseminated.
Unfortunately, the current educational efforts in AI in NP---while extremely valuable---are patchwork. 
They include summer schools, topical programs, workshops, and conferences.

A coherent inter-disciplinary approach is needed. Several mechanisms were discussed at the Workshop aiming at improving the situation.
\begin{description}
\item[University-wide AI courses] There is a need for inter-disciplinary AI courses involving Applied Mathematics,  Statistics, and Computer Science experts, as well as domain scientists. Online courses play important role, but the in-person approaches are superior. 
\item[Graduate Fellowships]
Establishing graduate NP/AI fellowships, similar to, e.g.,
\href{https://www.krellinst.org/csgf/about-doe-csgf}{DOE Computational Science Graduate Fellowship} or
\href{https://www.krellinst.org/ssgf/about-doe-nnsa-ssgf}{DOE NNSA Stewardship Science Graduate Fellowship},  would enable the development of a well-educated workforce in this area.
\item[Dual Ph.D. Programs]
Some universities  allow ``dual Ph.D.'' programs that allow individual students to work within two different graduate programs.  Students start graduate school in their primary department, and then enter such a program by arranging a secondary affiliation upon choice of a research project and advisor. Certificates in AI/ML are a less intensive but still beneficial approach to this problem.
\item[Educational Outreach Opportunities] 
AI practice is inherently interdisciplinary and an effort should be made to introduce the AI field to young physicists, computer and data scientists, mathematicians, and others in related fields as they choose their career paths. Conferences and Workshops play an important role in this cross-pollination.  For example, workshop organizers received a grant from the National Science Foundation that funded travel for 18 graduate- and undergraduate-students and early career professionals, most of whom indicated they would not have been able to attend the workshop without this support.  The pre-workshop hackathon provided students with an opportunity to creatively collaborate on a problem solving competition related to AI.

\end{description}

\section{The Level of AI Literacy} \label{Literacy}

\noindent 
The interest in the workshop was very good: as many as 184 scientists came to the meeting and many attended remotely. 
According to the data gathered by the Workshop's questionnaire, around 40\% participants are new to AI, 70\% would like to apply techniques from this workshop, and 40\% actively working on project using AI. These numbers well reflect the current situation: many nuclear physicists understand the potential benefits of AI, but there is a steep learning curve.

Considering the current efforts, more sophistication in using AI tools is needed. Indeed, 
majority of NP users  apply off-the-shelf tools; fewer  understand the AI glossary and make informed choices about the modern AI tools that suit their problem best. Even fewer practitioners are advanced users or innovators who
consider uncertainty quantification to be an essential part of the answer and/or
consider the full feedback between AI and physics problem (AI application is modified depending on the physics outcome).

In short, at this point, NP community at large does not fully grasp the depth of the  AI universe with the majority of work being carried out by 
 users often helped by enthusiastic undergraduate and graduate students. But the foundations are there: nuclear physicists have good technical background and they are used to {\it problem-driven approaches to tool selection}.  This helps in choosing the best/right tools for the problems. One has to remember, however, that the newest AI tools are almost always largely untested. It takes some experience to know which tools to use.  Simply understanding that this is true will help nuclear physicists avoid dangerous pitfalls. 
 
 How can the level of AI literacy be improved? As discussed in Sec.~\ref{Collaborations} the fastest route to an AI-educated community involves easy access to ML/AI/Data science experts. In the long-term, education of younger generation is essential. Several mechanism to improve the situation in this area are proposed in Sec.~\ref{Workforce}.

\clearpage\newpage

\bibliography{References}

\clearpage\newpage

\appendix

\section{Committees}

\noindent 
{\bf Local Organizing Committee}
\begin{itemize}
 \setlength\itemsep{3pt}    
\item 
Amber Boehnlein (Jefferson Lab)
\item Latifa Elouadrhiri (Jefferson Lab)
\item Robert McKeown (Jefferson Lab)
\item Yves Roblin (Jefferson Lab)
\end{itemize}

\noindent 
{\bf Advisory Committee}
\begin{itemize}
 \setlength\itemsep{3pt}  
\item {Robert McKeown, Jefferson Lab}
\item {Peter Alonzi, University of Virginia}
\item {Oliver Baker, Yale University}
\item {Jason Detwiler, University of Washington}
\item {Carl Gagliardi, Texas A\&M University}
\item {Sean Liddick, Michigan State University}
\item {Amy Lovell, Los Alamos National Lab}
\item {Bronson Messer, Oak Ridge National Lab}
\item {Curtis Meyer, Carnegie Mellon University}
\item {Zein-Eddine Meziani, Argonne National Lab}
\item {Jeff Nichols, Oak Ridge National Lab}
\item {Alan Poon, Lawrence Berkeley Lab}
\item {Alessandro Roggero, University of Washington}
\item {Phiala Shanahan, Massachusetts Institute of Technology}
\item {Torre Wenaus, Brookhaven National Lab}
\item {Ying Wu, Duke University}
\end{itemize}

\clearpage\newpage
\section{Working Groups and Conveners}
\begin{description}
\item[WG1. Lattice QCD and Other Quantum Field Theories] 
Paulo Bedaque (University of Maryland) and 
Kostas Orginos (William\&Mary/Jefferson Lab)

\item[WG2. Nuclear Structure Theory]
Witold Nazarewicz (Michigan State University) 
and Dean Lee (Michigan State University)

\item[WG3. Accelerator Science and Operations]            
Steven Lidia (Michigan State University) and
Malachi Schram (Pacific Northwest National Lab)

\item[WG4. Experimental Methods]                                  
Michelle Kuchera (Davidson College) and
Mario Cromaz (LBNL)

\item[WG5. Event Generation and Simulation]
Markus Diefenthaler (Jefferson Lab)

\item[WG6. Bayesian Inference for Quantum Correlation Functions]
Wally Melnitchouk (Jefferson Lab)

\end{description}

\end{document}